\documentstyle[10pt,epsf,dp_delphititle,axodraw]{dp_delphi}
\makeindex
\def\DpPaperGroup{PH-EP}
\def\DpPaperRef{2006-022}
\def\DpDate{19 June 2006/rev. 7 June 2007}
\def\DpAuthors{DELPHI Collaboration}
\def\DpSubmit{(Accepted by Eur. Phys. J. C)}
\def\DpTitle{\boldmath $Z\gamma^*$ production in $e^+ e^-$ interactions
at $\sqrt{s} = 183 - 209$ GeV}
\def\DpComment{}
\def\DpEMail{}

\newcommand{\ba}{\begin{array}}
\newcommand{\ea}{\end{array}}
\newcommand{\bc}{\begin{center}}
\newcommand{\ec}{\end{center}}
\newcommand{\bt}{\begin{tabular}}
\newcommand{\et}{\end{tabular}}
\newcommand{\beq}{\begin{eqnarray}}
\newcommand{\eeq}{\end{eqnarray}}
\newcommand{\bes}{\begin{eqnarray*}}
\newcommand{\ees}{\end{eqnarray*}}
\newcommand \llll {\ifmmode \ l^+l^-l^{'+}l^{'-}\else $l^+l^-l^{'+}l^{'-}$\fi}
\newcommand \eeee {\ifmmode \ e^+e^-e^+e^-\else $e^+e^-e^+e^-$\fi}
\newcommand \eemm {\ifmmode \ e^+e^-\mu^+\mu^-\else $e^+e^-\mu^+\mu^-$\fi}
\newcommand \eepp {\ifmmode \ e^+e^-\pi^+\pi^-\else $e^+e^-\pi^+\pi^-$\fi}
\newcommand \eett {\ifmmode \ e^+e^-\tau^+\tau^-\else $e^+e^-\tau^+\tau^-$\fi}
\newcommand \mmmm {\ifmmode \ \mu^+\mu^-\mu^+\mu^-\else
$\mu^+\mu^-\mu^+\mu^-$\fi}
\newcommand \mmtt {\ifmmode \ \mu^+\mu^-\tau^+\tau^-\else
$\mu^+\mu^-\tau^+\tau^-$\fi}
\newcommand \mmpp {\ifmmode \ \mu^+\mu^-\pi^+\pi^-\else
$\mu^+\mu^-\pi^+\pi^-$\fi}
\newcommand \tttt {\ifmmode \ \tau^+\tau^-\tau^+\tau^-\else
$\tau^+\tau^-\tau^+\tau^-$\fi}
\newcommand \pipi {\ifmmode \ \pi^+\pi^-\pi^+\pi^-\else
$\pi^+\pi^-\pi^+\pi^-$\fi}
\newcommand \eell {\ifmmode \ e^+e^-l^+l^-\else $e^+e^-l^+l^-$\fi}
\newcommand{\dgree}   {\mbox{$ ^\circ                                      $}}

\begin{document}
\makeatletter
\newcount\@tempcntc
\def\@citex[#1]#2{\if@filesw\immediate\write\@auxout{\string\citation{#2}}\fi
  \@tempcnta\z@\@tempcntb\m@ne\def\@citea{}\@cite{\@for\@citeb:=#2\do
    {\@ifundefined
       {b@\@citeb}{\@citeo\@tempcntb\m@ne\@citea\def\@citea{,}{\bf ?}\@warning
       {Citation `\@citeb' on page \thepage \space undefined}}%
    {\setbox\z@\hbox{\global\@tempcntc0\csname b@\@citeb\endcsname\relax}%
     \ifnum\@tempcntc=\z@ \@citeo\@tempcntb\m@ne
       \@citea\def\@citea{,}\hbox{\csname b@\@citeb\endcsname}%
     \else
      \advance\@tempcntb\@ne
      \ifnum\@tempcntb=\@tempcntc
      \else\advance\@tempcntb\m@ne\@citeo
      \@tempcnta\@tempcntc\@tempcntb\@tempcntc\fi\fi}}\@citeo}{#1}}
\def\@citeo{\ifnum\@tempcnta>\@tempcntb\else\@citea\def\@citea{,}%
  \ifnum\@tempcnta=\@tempcntb\the\@tempcnta\else
   {\advance\@tempcnta\@ne\ifnum\@tempcnta=\@tempcntb \else \def\@citea{--}\fi
    \advance\@tempcnta\m@ne\the\@tempcnta\@citea\the\@tempcntb}\fi\fi}
 
\makeatother
\begin{titlepage}
\pagenumbering{roman}
\CERNpreprint{\DpPaperGroup}{\DpPaperRef} 
\date{{\small\DpDate}} 
\title{\DpTitle} 
\address{\DpAuthors} 
\begin{shortabs} 
\noindent
%
\noindent

Measurements of $Z \gamma^*$ production are presented using  data collected by the DELPHI  detector at centre-of-mass energies ranging from 183~to 209~GeV,  corresponding to an integrated luminosity of 
about 667~pb$^{-1}$. The measurements cover a wide range of the possible final state four-fermion
configurations: hadronic and leptonic ($e^+ e^- q \bar q$, $\mu^+ \mu^- q \bar q$, $q \bar q \nu \bar \nu$), 
fully leptonic (\llll) and fully hadronic final states  ($q \bar q q \bar q$,  with a low mass $q \bar q$  pair). 
Measurements of the $Z \gamma^*$ cross-section for the  various final states have been compared with the 
Standard  Model expectations and found to be consistent within the errors. In addition, a total cross-section measurement of the \llll\ cross-section is reported, and found to be in agreement with the prediction of the Standard Model.

\end{shortabs}
\vfill
\begin{center}
\DpSubmit \ \\ 
\DpComment \ \\
\DpEMail \ \\
\end{center}
\vfill
\clearpage
\headsep 10.0pt
\addtolength{\textheight}{10mm}
\addtolength{\footskip}{-5mm}
\begingroup
%
\newcommand{\DpName}[2]{\hbox{#1$^{\ref{#2}}$},\hfill}
\newcommand{\DpNameTwo}[3]{\hbox{#1$^{\ref{#2},\ref{#3}}$},\hfill}
\newcommand{\DpNameThree}[4]{\hbox{#1$^{\ref{#2},\ref{#3},\ref{#4}}$},\hfill}
\newskip\Bigfill \Bigfill = 0pt plus 1000fill
\newcommand{\DpNameLast}[2]{\hbox{#1$^{\ref{#2}}$}\hspace{\Bigfill}}
%
\footnotesize
\noindent
\DpName{J.Abdallah}{LPNHE}
\DpName{P.Abreu}{LIP}
\DpName{W.Adam}{VIENNA}
\DpName{P.Adzic}{DEMOKRITOS}
\DpName{T.Albrecht}{KARLSRUHE}
\DpName{R.Alemany-Fernandez}{CERN}
\DpName{T.Allmendinger}{KARLSRUHE}
\DpName{P.P.Allport}{LIVERPOOL}
\DpName{U.Amaldi}{MILANO2}
\DpName{N.Amapane}{TORINO}
% ...
\normalsize
\endgroup
\newpage
\titlefoot{Department of Physics and Astronomy, Iowa State
     University, Ames IA 50011-3160, USA
    \label{AMES}}
\titlefoot{IIHE, ULB-VUB,
     Pleinlaan 2, B-1050 Brussels, Belgium
    \label{BRUSSELS}}
\titlefoot{Physics Laboratory, University of Athens, Solonos Str.
     104, GR-10680 Athens, Greece
    \label{ATHENS}}
\titlefoot{Department of Physics, University of Bergen,
     All\'egaten 55, NO-5007 Bergen, Norway
    \label{BERGEN}}
\titlefoot{Dipartimento di Fisica, Universit\`a di Bologna and INFN,
     Via Irnerio 46, IT-40126 Bologna, Italy
    \label{BOLOGNA}}
\titlefoot{Centro Brasileiro de Pesquisas F\'{\i}sicas, rua Xavier Sigaud 150,
     BR-22290 Rio de Janeiro, Brazil
    \label{BRASIL-CBPF}}
\titlefoot{Inst. de F\'{\i}sica, Univ. Estadual do Rio de Janeiro,
     rua S\~{a}o Francisco Xavier 524, Rio de Janeiro, Brazil
    \label{BRASIL-IFUERJ}}
\titlefoot{Coll\`ege de France, Lab. de Physique Corpusculaire, IN2P3-CNRS,
     FR-75231 Paris Cedex 05, France
    \label{CDF}}
\titlefoot{CERN, CH-1211 Geneva 23, Switzerland
    \label{CERN}}
\titlefoot{Institut de Recherches Subatomiques, IN2P3 - CNRS/ULP - BP20,
     FR-67037 Strasbourg Cedex, France
    \label{CRN}}
\titlefoot{Now at DESY-Zeuthen, Platanenallee 6, D-15735 Zeuthen, Germany
    \label{DESY}}
\titlefoot{Institute of Nuclear Physics, N.C.S.R. Demokritos,
     P.O. Box 60228, GR-15310 Athens, Greece
    \label{DEMOKRITOS}}
\titlefoot{FZU, Inst. of Phys. of the C.A.S. High Energy Physics Division,
     Na Slovance 2, CZ-182 21, Praha 8, Czech Republic
    \label{FZU}}
\titlefoot{Dipartimento di Fisica, Universit\`a di Genova and INFN,
     Via Dodecaneso 33, IT-16146 Genova, Italy
    \label{GENOVA}}
\titlefoot{Institut des Sciences Nucl\'eaires, IN2P3-CNRS, Universit\'e
     de Grenoble 1, FR-38026 Grenoble Cedex, France
    \label{GRENOBLE}}
\titlefoot{Helsinki Institute of Physics and Department of Physical Sciences,
     P.O. Box 64, FIN-00014 University of Helsinki, 
     \indent~~Finland
    \label{HELSINKI}}
\titlefoot{Joint Institute for Nuclear Research, Dubna, Head Post
     Office, P.O. Box 79, RU-101 000 Moscow, Russian Federation
    \label{JINR}}
\titlefoot{Institut f\"ur Experimentelle Kernphysik,
     Universit\"at Karlsruhe, Postfach 6980, DE-76128 Karlsruhe,
     Germany
    \label{KARLSRUHE}}
\titlefoot{Institute of Nuclear Physics PAN,Ul. Radzikowskiego 152,
     PL-31142 Krakow, Poland
    \label{KRAKOW1}}
\titlefoot{Faculty of Physics and Nuclear Techniques, University of Mining
     and Metallurgy, PL-30055 Krakow, Poland
    \label{KRAKOW2}}
\titlefoot{Universit\'e de Paris-Sud, Lab. de l'Acc\'el\'erateur
     Lin\'eaire, IN2P3-CNRS, B\^{a}t. 200, FR-91405 Orsay Cedex, France
    \label{LAL}}
\titlefoot{School of Physics and Chemistry, University of Lancaster,
     Lancaster LA1 4YB, UK
    \label{LANCASTER}}
\titlefoot{LIP, IST, FCUL - Av. Elias Garcia, 14-$1^{o}$,
     PT-1000 Lisboa Codex, Portugal
    \label{LIP}}
\titlefoot{Department of Physics, University of Liverpool, P.O.
     Box 147, Liverpool L69 3BX, UK
    \label{LIVERPOOL}}
\titlefoot{Dept. of Physics and Astronomy, Kelvin Building,
     University of Glasgow, Glasgow G12 8QQ
    \label{GLASGOW}}
\titlefoot{LPNHE, IN2P3-CNRS, Univ.~Paris VI et VII, Tour 33 (RdC),
     4 place Jussieu, FR-75252 Paris Cedex 05, France
    \label{LPNHE}}
\titlefoot{Department of Physics, University of Lund,
     S\"olvegatan 14, SE-223 63 Lund, Sweden
    \label{LUND}}
\titlefoot{Universit\'e Claude Bernard de Lyon, IPNL, IN2P3-CNRS,
     FR-69622 Villeurbanne Cedex, France
    \label{LYON}}
\titlefoot{Dipartimento di Fisica, Universit\`a di Milano and INFN-MILANO,
     Via Celoria 16, IT-20133 Milan, Italy
    \label{MILANO}}
\titlefoot{Dipartimento di Fisica, Univ. di Milano-Bicocca and
     INFN-MILANO, Piazza della Scienza 3, IT-20126 Milan, Italy
    \label{MILANO2}}
\titlefoot{IPNP of MFF, Charles Univ., Areal MFF,
     V Holesovickach 2, CZ-180 00, Praha 8, Czech Republic
    \label{NC}}
\titlefoot{NIKHEF, Postbus 41882, NL-1009 DB
     Amsterdam, The Netherlands
    \label{NIKHEF}}
\titlefoot{National Technical University, Physics Department,
     Zografou Campus, GR-15773 Athens, Greece
    \label{NTU-ATHENS}}
\titlefoot{Physics Department, University of Oslo, Blindern,
     NO-0316 Oslo, Norway
    \label{OSLO}}
\titlefoot{Dpto. Fisica, Univ. Oviedo, Avda. Calvo Sotelo
     s/n, ES-33007 Oviedo, Spain
    \label{OVIEDO}}
\titlefoot{Department of Physics, University of Oxford,
     Keble Road, Oxford OX1 3RH, UK
    \label{OXFORD}}
\titlefoot{Dipartimento di Fisica, Universit\`a di Padova and
     INFN, Via Marzolo 8, IT-35131 Padua, Italy
    \label{PADOVA}}
\titlefoot{Rutherford Appleton Laboratory, Chilton, Didcot
     OX11 OQX, UK
    \label{RAL}}
\titlefoot{Dipartimento di Fisica, Universit\`a di Roma II and
     INFN, Tor Vergata, IT-00173 Rome, Italy
    \label{ROMA2}}
\titlefoot{Dipartimento di Fisica, Universit\`a di Roma III and
     INFN, Via della Vasca Navale 84, IT-00146 Rome, Italy
    \label{ROMA3}}
\titlefoot{DAPNIA/Service de Physique des Particules,
     CEA-Saclay, FR-91191 Gif-sur-Yvette Cedex, France
    \label{SACLAY}}
\titlefoot{Instituto de Fisica de Cantabria (CSIC-UC), Avda.
     los Castros s/n, ES-39006 Santander, Spain
    \label{SANTANDER}}
\titlefoot{Inst. for High Energy Physics, Serpukov
     P.O. Box 35, Protvino, (Moscow Region), Russian Federation
    \label{SERPUKHOV}}
\titlefoot{J. Stefan Institute, Jamova 39, SI-1000 Ljubljana, Slovenia
    \label{SLOVENIJA1}}
\titlefoot{Laboratory for Astroparticle Physics,
     University of Nova Gorica, Kostanjeviska 16a, SI-5000 Nova Gorica, Slovenia
    \label{SLOVENIJA2}}
\titlefoot{Department of Physics, University of Ljubljana,
     SI-1000 Ljubljana, Slovenia
    \label{SLOVENIJA3}}
\titlefoot{Fysikum, Stockholm University,
     Box 6730, SE-113 85 Stockholm, Sweden
    \label{STOCKHOLM}}
\titlefoot{Dipartimento di Fisica Sperimentale, Universit\`a di
     Torino and INFN, Via P. Giuria 1, IT-10125 Turin, Italy
    \label{TORINO}}
\titlefoot{INFN,Sezione di Torino and Dipartimento di Fisica Teorica,
     Universit\`a di Torino, Via Giuria 1,
     IT-10125 Turin, Italy
    \label{TORINOTH}}
\titlefoot{Dipartimento di Fisica, Universit\`a di Trieste and
     INFN, Via A. Valerio 2, IT-34127 Trieste, Italy
    \label{TRIESTE}}
\titlefoot{Istituto di Fisica, Universit\`a di Udine and INFN,
     IT-33100 Udine, Italy
    \label{UDINE}}
\titlefoot{Univ. Federal do Rio de Janeiro, C.P. 68528
     Cidade Univ., Ilha do Fund\~ao
     BR-21945-970 Rio de Janeiro, Brazil
    \label{UFRJ}}
\titlefoot{Department of Radiation Sciences, University of
     Uppsala, P.O. Box 535, SE-751 21 Uppsala, Sweden
    \label{UPPSALA}}
\titlefoot{IFIC, Valencia-CSIC, and D.F.A.M.N., U. de Valencia,
     Avda. Dr. Moliner 50, ES-46100 Burjassot (Valencia), Spain
    \label{VALENCIA}}
\titlefoot{Institut f\"ur Hochenergiephysik, \"Osterr. Akad.
     d. Wissensch., Nikolsdorfergasse 18, AT-1050 Vienna, Austria
    \label{VIENNA}}
\titlefoot{Inst. Nuclear Studies and University of Warsaw, Ul.
     Hoza 69, PL-00681 Warsaw, Poland
    \label{WARSZAWA}}
\titlefoot{Now at University of Warwick, Coventry CV4 7AL, UK
    \label{WARWICK}}
\titlefoot{Fachbereich Physik, University of Wuppertal, Postfach
     100 127, DE-42097 Wuppertal, Germany \\
\noindent
{$^\dagger$~deceased}
    \label{WUPPERTAL}}
\addtolength{\textheight}{-10mm}
\addtolength{\footskip}{5mm}
\clearpage
\headsep 30.0pt
%
\end{titlepage}
%
\pagenumbering{arabic} 
\setcounter{footnote}{0} %
\large


%

\section{Introduction \label{sec:INTRO}}

The study of four-fermion processes in $e^+ e^-$ interactions becomes
increasingly important as the centre-of-mass energy and the corresponding
luminosity increase. The main goal of such studies is to verify the Standard
Model predictions and to look for, or to set limits on, possible contributions
arising from mechanisms beyond the Standard Model: for instance, anomalous triple gauge
couplings \cite{hagi}  can usually give contributions to
four-fermion final states. Moreover, such processes  form an important
background to new particle searches, such as those for charginos, 
neutralinos or non-standard Higgs bosons, and  deviations from the Standard Model expectations
would be a signal of new  physics.  LEP has provided a unique opportunity to
study four-fermion interactions at several energies. On-shell pair production of
$W$ \cite{WW} and $Z$ \cite{ZZdelphi,ZZ}  bosons has been studied extensively. The focus
of this paper is the measurement of the cross-section of neutral current
processes with a $Z$ and an off-shell photon  ($Z\gamma^*$ in the following).
To this end, several channels were  studied:    $l^+ l^- q \bar q ~(l \equiv e ,
\mu )$, $q \bar{q} \nu \bar \nu$, \llll\  ($l,l^{'} \equiv e, \mu , \tau$)  and
$q \bar q q \bar q$ (with a low mass $q \bar q$ pair).  In addition, for \llll\
final states, a measurement of the total neutral current cross-section has
been performed.

\begin{figure}[tbh]
  \begin{center}
    \mbox{\epsfysize=15cm\epsffile{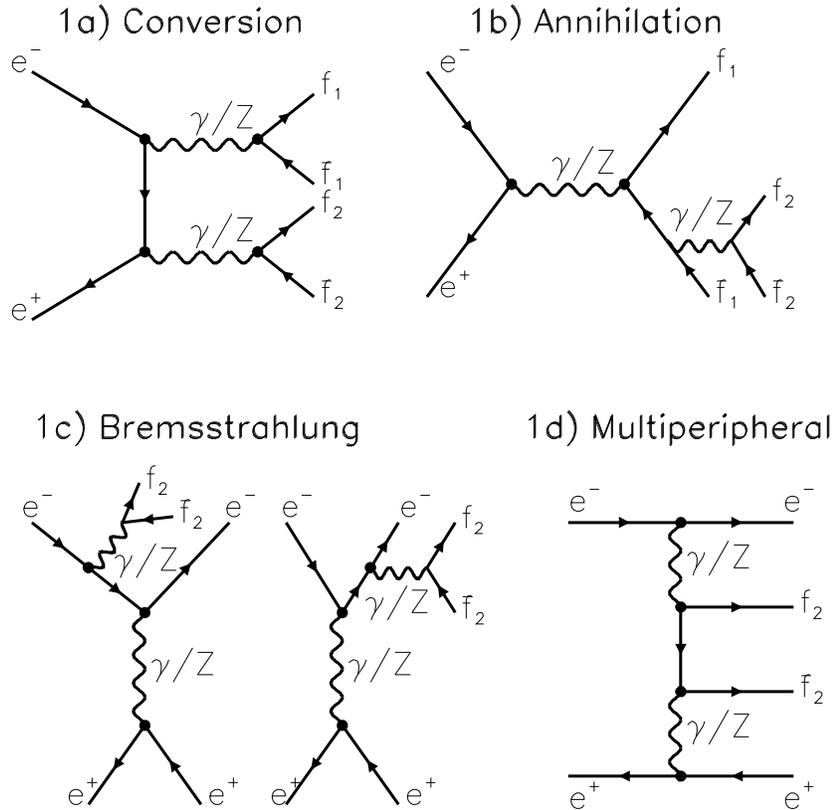}}
\caption{The Feynman diagrams for four-fermion neutral current production
  in $e^+e^-$ collisions.}
 \label{fig:zgconv}
  \end{center}
\end{figure}

Figure~\ref{fig:zgconv} shows the main Feynman diagrams for four-fermion neutral current production in $e^+e^-$ collisions. When there are no electrons in the final state these processes are dominated by the conversion processes shown in figure~\ref{fig:zgconv}a.  This graph represents eight different diagrams, usually referred to as the NC08 diagrams:  two of them (usually referred to as NC02) lead to $ZZ$ production,  two to $\gamma^* \gamma^*$ and four to $Z\gamma^*$. These four $Z\gamma^*$ conversion diagrams are sketched in figure~\ref{fig:zg}, and the square of their summed amplitudes is used in the definition of the signal to be measured in this paper, as explained below. A prominent feature of the graphs in figure~\ref{fig:zg} is the very different scale of the momentum transfer at the $Z f {\bar f}$ and $\gamma f {\bar f}$ vertices, an issue which must be properly addressed by the simulation programs (see section~\ref{sec:DETEC}). For final states with electrons, other processes, such as $t$-channel $\gamma$ exchange accompanied by  $Z^*/\gamma^*$-strahlung (figure~\ref{fig:zgconv}c) and multiperipheral production (figure~\ref{fig:zgconv}d), contribute significantly. In particular, the processes originating from the bremsstrahlung diagram  (\ref{fig:zgconv}c), usually referred to as $Zee$ and $\gamma^*ee$, represent an important background to the measurements with electrons presented in this paper (see sections \ref{sec:LLQQ}~and~\ref{sec:4LEPT}). Interference effects of these processes with those originating from $Z\gamma^*$ can also be important and have to be taken into account.

The $Z\gamma^*$ production cross-section depends weakly on the centre-of-mass energy, but strongly on the mass of the virtual photon. For real photon production, $e^+e^- \rightarrow Z\gamma$, the cross-section reaches values above $100$~pb, while in the kinematic region of $Z\gamma^*$ production considered in this paper, its value is generally in the region of a fraction of a picobarn. Furthermore, in the $Z\gamma^*$ production processes, particles coming from the conversion of low mass $\gamma^*$s into hadrons or leptons are preferentially produced at very small angles with respect to the beam direction. A measurement of this cross-section has thus to be performed for a specific selection of the $\gamma^*$ mass and production polar angle. 

Data collected by the DELPHI experiment in 1997-2000 at centre-of-mass energies from 183 to 209~GeV were used, corresponding to an integrated luminosity of about 667~pb$^{-1}$. Results for each channel are given in the form of a
comparison of the predicted numbers of selected events with those found in data. Combination of channel results into an overall $Z\gamma^*$ cross-section is then performed. The resulting measurement is compared to the Standard Model expectation.  The results presented here complement and augment those reported in previous studies of neutral current four-fermion production at LEP~\cite{otherzg}.

This paper is organised as follows. Two definitions of the $Z\gamma^*$ signal are given in section~\ref{sec:signal}, one, the ``Matrix Element definition", according to the Feynman diagrams contributing to $Z\gamma^*$ production, the other, the ``LEP definition", using invariant mass cuts. Short descriptions of the detector, of the available data sets and of the 
simulation programs  used in the analyses are given in section~\ref{sec:DETEC}. The subsequent sections provide descriptions of the analyses used for the first signal definition for each of the 
channels studied:  
$l^+l^-q \bar q$ (section~\ref{sec:LLQQ}), $q \bar q \nu \bar \nu$  
(section~\ref{sec:VVQQ}), \llll\ (section~\ref{sec:4LEPT}, where a total 
cross-section measurement is also presented), and  $q \bar q q \bar q$ 
(section~\ref{sec:QQQQ}). The results using the Matrix Element signal definition 
are presented in section~\ref{sec:RESULTS}, while the analyses and results 
using the LEP signal definition are described in 
section~\ref{sec:RESULTS_LEP}.   Conclusions are given in section~\ref{sec:CONCL}. 

\section{Signal definition \label{sec:signal}}
\begin{figure}[tbh]
  \begin{center}
    \mbox{\epsfysize=10cm\epsffile{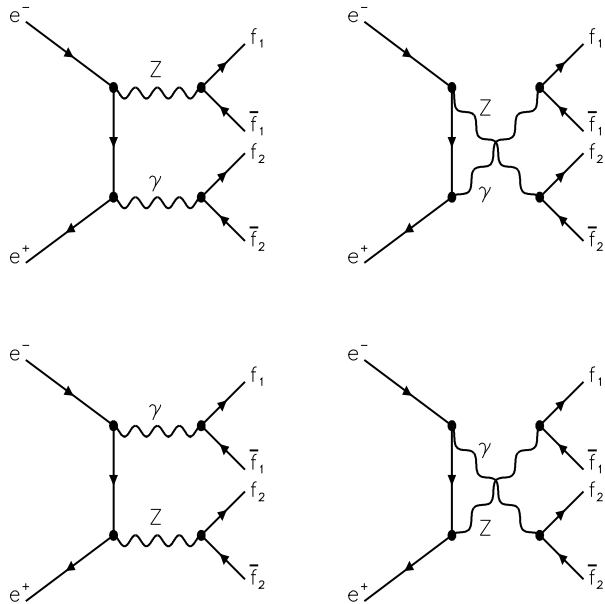}}
\caption{Neutral current conversion diagrams for the $Z \gamma^*$ process.}
 \label{fig:zg}
  \end{center}
\end{figure}

Two different signal definitions were adopted in the analyses presented in this paper:
\begin{itemize}
\item  
{\bf The Matrix Element definition:}
For each of the final states considered, the signal was first defined by applying the  following kinematic selection on all charged fermions at generator level:
 $$ |\cos \theta_{f^\pm}|<0.98~, $$
where $\theta_{f^\pm}$ is the polar angle of the charged fermion with respect to the beam axis. Events with one or more charged fermions not fulfilling these selections were considered as background.  Then, for the surviving events, the signal was defined as the $Z\gamma^*$ contribution coming from the four conversion diagrams shown  in
figure~\ref{fig:zg}. This was achieved by weighting the events in the selected generator-level sample  by the quantity 
 $$\frac{| {\cal M}_{Z \gamma^*}|^2}{|{\cal M}_{all}|^2}~,$$
where ${\cal M}_{Z \gamma^*}$ and ${\cal M}_{all}$ are the matrix elements for $Z\gamma^*$ and for all the graphs in figure~\ref{fig:zgconv}, respectively. Analogously, using the same weighting technique, the components obtained by
weighting events by the quantities  
$\frac{|{\cal M}_{all-Z \gamma^*}|^2}{|{\cal M}_{all}|^2}$  and  $1-\frac{|{\cal M}_{Z\gamma^*}|^2}{|{\cal M}_{all}|^2}-\frac{|{\cal M}_{all-Z \gamma^*}|^2}{|{\cal M}_{all}|^2}$ 
were considered as background: these components represent, respectively,  the contributions arising from non-$Z\gamma^*$ four-fermion processes (including $ZZ$ and $\gamma^* \gamma^*$, which are also produced via  
conversion diagrams) and from the interference effects between  $Z\gamma^*$ and non-$Z\gamma^*$ graphs. Expected rates were thus computed using generated events weighted by the appropriate ratio. Efficiencies were defined from the simulated event samples as the ratio of selected weighted events over all weighted events.

\item {\bf The LEP definition:}
The second definition was agreed between the LEP Collaborations in order
to combine results in a meaningful way. It is based on invariant mass cuts at
generator level and explicitly avoids the difficult regions of low di-fermion 
masses.  Depending on the final state, the following cuts were applied on
invariant masses  of fermion pairs and, where relevant, on lepton production polar angles:
$M_{q \bar q}>10$~GeV/$c^2$, $M_{l^+ l^-}>5$~GeV/$c^2$, $|\cos
\theta_{l^\pm}|<0.95$. Furthermore, it was required that only one fermion pair
in the event had an invariant mass, $M_{f^+f^-}$, satisfying  
$|M_{f^+ f^-}-M_Z|< 2 \Gamma_Z$, where $M_Z$ and $\Gamma_Z$ are the 
nominal mass and width of the $Z$ boson.  
Only the three dominant channels in the final result combination  
($\mu^+ \mu^- q \bar q$, $e^+ e^- q \bar q$ and $q \bar q \nu \bar \nu$) 
were analysed using the LEP signal definition. 
\end{itemize} 

\noindent In the rest of this paper, when not explicitly stated, it is implied 
that the Matrix Element  signal definition is being used.

\section{Detector description and simulation \label{sec:DETEC}}

A detailed description of the DELPHI detector and a review of its performance
can be found in~\cite{delphi1,delphi2}. For LEP2 operations,  the vertex
detector was upgraded~\cite{VD}, and a set of scintillation counters was added
to veto photons in blind regions of the electromagnetic calorimetry, at polar
angles around $\theta=40^\circ$, $\theta=90^\circ$ and $\theta=140^\circ$. 

The integrated luminosity of 666.7~pb$^{-1}$ collected by the DELPHI detector at  centre-of-mass energies
from  182.7~to 209~GeV was used in the analysis. The luminosities collected
at various centre-of-mass energies are shown in table~\ref{tab:lumi}.

\begin{table}[htb]
\begin{center}
\begin{tabular}{|c|c|c|} 
\hline
Year       & $\sqrt s$     & Integrated              \\
           & $[$GeV$]$     & luminosity [pb$^{-1}$]  \\ 
\hline
1997       & 182.7         & 55.0                   \\
1998       & 188.6         & 158.1                  \\
1999       & 191.6         & 25.0                   \\
1999       & 195.5         & 77.0                   \\
1999       & 199.5         & 82.0                   \\
1999       & 201.6         & 41.0                   \\
2000       & 205.0         & 81.3                   \\ 
2000       & 206.5         & 147.3                  \\
\hline
Total       & 197.1          & 666.7                  \\
\hline
\end{tabular}
\end{center}
\caption{\label{tab:lumi} 
Luminosity-weighted centre-of-mass energies and integrated luminosities of the data analysed.
During the year 2000, the energies reached were in the range 202-209~GeV and 
clustered mainly around 205 and 206.5~GeV. 
}
\end{table}

During the year 2000, one sector (1/12) of the main tracking device, the Time
Projection Chamber (TPC), was inactive from the  beginning of September to the
end of data taking, which corresponded to about a quarter of the 2000 data
sample. The effect was taken into account in the detector simulation and the 
corresponding small change of analysis sensitivity for this period was
considered in the extraction of the cross-sections.

Simulated events were produced with the DELPHI simulation program 
DELSIM~\cite{delphi2} and then passed through the same reconstruction chain as
the data.  The generation of processes leading to four-fermion final  states,
mediated by charged and neutral currents, was done with  
WPHACT~\cite{wphact,4f},  interfaced to the  PYTHIA~\cite{pythia} fragmentation
and hadronisation model. For the charged current part, WPHACT incorporates the
$O(\alpha)$ Double Pole Approximation~\cite{oa1,oa2} radiative corrections  to
the doubly resonant $WW$ production diagrams via a weighting technique, 
with the matrix elements provided by the YFSWW generator~\cite{yfsww}. At a general
level, WPHACT performs fully massive calculations all over the phase space,
includes higher-order corrections and uses the package QEDPS~\cite{qedps} for
initial state radiation. Two additional features, particularly relevant  for the
analyses described in this paper, were implemented in WPHACT: the study of the
most suitable scale to use for $\alpha_{QED}$ at the $\gamma^*$ vertices of the
diagrams in figure~\ref{fig:zg}, and the treatment of the hadronisation of low
mass virtual photons. The first of these problems was solved in WPHACT by
implementing the running of  $\alpha_{QED}$ at the level of the event
generation, thus using the value of the coupling constant corresponding to the
mass of the photon propagator at the $\gamma^*$ vertices.  The second problem
was addressed by interfacing WPHACT  with a special package~\cite{maarten} for
the specific treatment of the hadronisation of low mass $q \bar q$ systems. 
This package provides a description of the hadronisation from the $\gamma^*
\rightarrow q \bar q$ process in the mass region below 2 GeV/$c^2$ both due to
the presence of hadronic resonances (with subsequent decays described by 
PYTHIA) and in the continuum, based on experimental $e^+ e^-$ data at low energy. This
is particularly important for the $q \bar{q} \nu \bar \nu$ (section
\ref{sec:VVQQ})  and $q \bar q q \bar q$ (section \ref{sec:QQQQ}) analyses, 
which explore the low mass $q \bar q$ region.  Phase space cuts are applied in
WPHACT and can be found in table~2 of~\cite{4f}.  The study of the backgrounds
due to $q \bar q (\gamma)$,  $\mu^+ \mu^-(\gamma)$ and $\tau^+ \tau^-(\gamma)$
production was made using the KK2f~\cite{kk2f} model; $e^+ e^-(\gamma)$ events
were simulated with BHWIDE~\cite{bhwide}. Two-photon interactions were 
generated using WPHACT for the regions in which the multiperipheral contribution is not
dominant and using BDK~\cite{bdk} for the pure two-photon region; 
PYTHIA~6.143 was used to model two-photon processes with single and doubly
resolved photons.

\section {Study of the $l^+ l^- q \bar{q}$ final state}
\label{sec:LLQQ}

In this section the analysis of the final state containing jets and a pair of
identified muons or electrons is described. The two final state leptons in the
process  $e^+ e^- \rightarrow l^+ l^- q \bar{q}$ are usually well isolated
from all other particles. This property can be used to select such events with
high efficiency in both the muon and electron channels. Events with $\tau^+
\tau^-$ pairs have not been considered here. This part of the analysis follows
very closely the one performed in~\cite{ZZdelphi}, where an identical final
state was studied. 

A loose hadronic preselection was first applied, requiring that the events have at least 7 charged particles and a charged energy above 0.30$\sqrt{s}$. To suppress the radiative return  to the $Z$ (final state on-shell $Z$ production with the emission of a hard initial state radiation (ISR) photon) the event was rejected if a photon with energy more than 60~GeV was found.  The selection procedures were then carried out in a similar way for the $\mu^+ \mu^- q \bar{q}$ and $e^+ e^- q \bar{q}$ channels.  In order to maximise the lepton identification efficiency, any charged particle with momentum exceeding 5~GeV/$c$ was considered as a possible lepton candidate around which nearby photons, if present, could be clustered. This was found to be necessary to improve the energy evaluation in the presence of final state radiation from electrons. In the case of  the $e^+ e^- q \bar{q}$ channel, photons with energy between 20~GeV and 60~GeV were also considered as electron candidates in order to recover events in which the electron track was not reconstructed. For both electrons and muons, ``strong" and ``soft" identification criteria were then defined. Muons were considered as strongly identified if selected by the standard DELPHI muon identification package~\cite{delphi2}, based mainly on finding associated hits in the muon chambers. For soft muon identification, only kinematic and calorimetric criteria were used. Electrons were considered as strongly identified when the energy deposited in the electromagnetic calorimeter exceeded 60\% of the cluster energy or 15~GeV, whichever was greater, and when the energy deposited in the hadron calorimeter was less than a specified limit. For soft electron identification, only requirements on the momentum of the charged particle in the cluster and on the energy deposited in the hadron calorimeter were imposed. Electron candidates originating from applying the clustering procedure around a photon were considered as softly identified. Events with at least two lepton candidates of the same flavour and opposite charge were then selected.\footnote{The requirement of having  leptons of opposite charge was dropped in the case of candidate electrons originally identified as photons, for which no charge information is available.} All particles except the lepton candidates were clustered into two jets and a kinematic fit requiring four-momentum conservation was applied, after appropriately adjusting the errors on lepton energies in cases where photons had been added by the clustering procedure.

At least one of the two lepton candidates was required to satisfy strong lepton
identification criteria, while softer requirements were specified for the
second.

Two discriminating variables were then used for final event selection:
$P_t^{min}$, the lesser of the transverse momenta of the lepton candidates with
respect to the nearest jet, and the $\chi^2$ per degree of freedom of the
kinematic fit. Cuts on these variables were applied, with values depending on
the final state and on the quality of the lepton 
identification (see~\cite{ZZdelphi}). 

\subsection{Results \label{sec:LLQQRESULTS}}

\begin{figure}[tbh]
  \begin{center}
    \mbox{\epsfxsize=12.95cm\epsffile{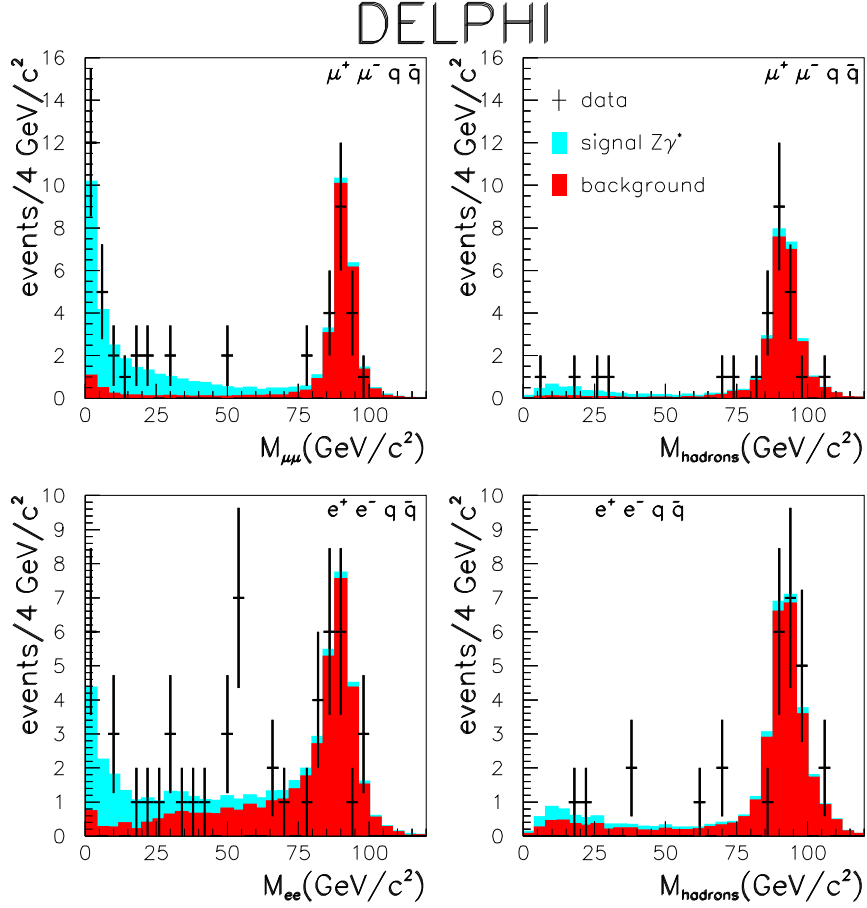}}
\caption{ 
$l^+ l^- q \bar{q}$ final state: Distributions of the mass of one fermion pair
when the mass of the second is within 15~GeV/$c^2$ of $M_Z$. The two lower 
plots are for the $e^+e^- q \bar{q}$ channel and the two upper plots for the
$\mu^+\mu^- q \bar{q}$  channel. The points are the data summed over all 
energy points, the dark (red) histogram is the distribution of the background predicted by the Standard Model, and the light (light blue)  histogram is the predicted distribution of the $Z \gamma^*$ signal.}
    \label{fig:llqq1}
  \end{center}
\end{figure}

The distribution of the reconstructed mass of one fermion pair when the mass of the second one is within 15~GeV/$c^2$ of the nominal $Z$ mass is compared with the predictions of the Standard Model in figure~\ref{fig:llqq1}, separately for $\mu^+ \mu^- q \bar q$ and $e^+ e^- q \bar q$ events. In the  $\mu^+ \mu^- q \bar q$ channel the $Z \gamma^*$ contribution is clearly separated from the background component and is mostly concentrated in the region of the decay $Z \rightarrow q \bar q$, as expected. In the $e^+ e^- q \bar q$ case, the distribution is flatter, indicating the presence of non-resonant diagrams. In both channels there is good overall agreement between the observed and predicted numbers of events. In the $e^+ e^- q \bar q$ channel an accumulation of events is observed in the invariant mass distribution of the $e^+ e^-$ pair in the region between 50 and 60~GeV/$c^2$, with 7 events observed where 2.4 are expected. Various studies and comparisons with results of the other LEP experiments were performed, leading to the conclusion that this excess is most probably due to a statistical fluctuation.

\begin{figure}[h]
  \begin{center}
   \mbox{\epsfysize=13cm\epsffile{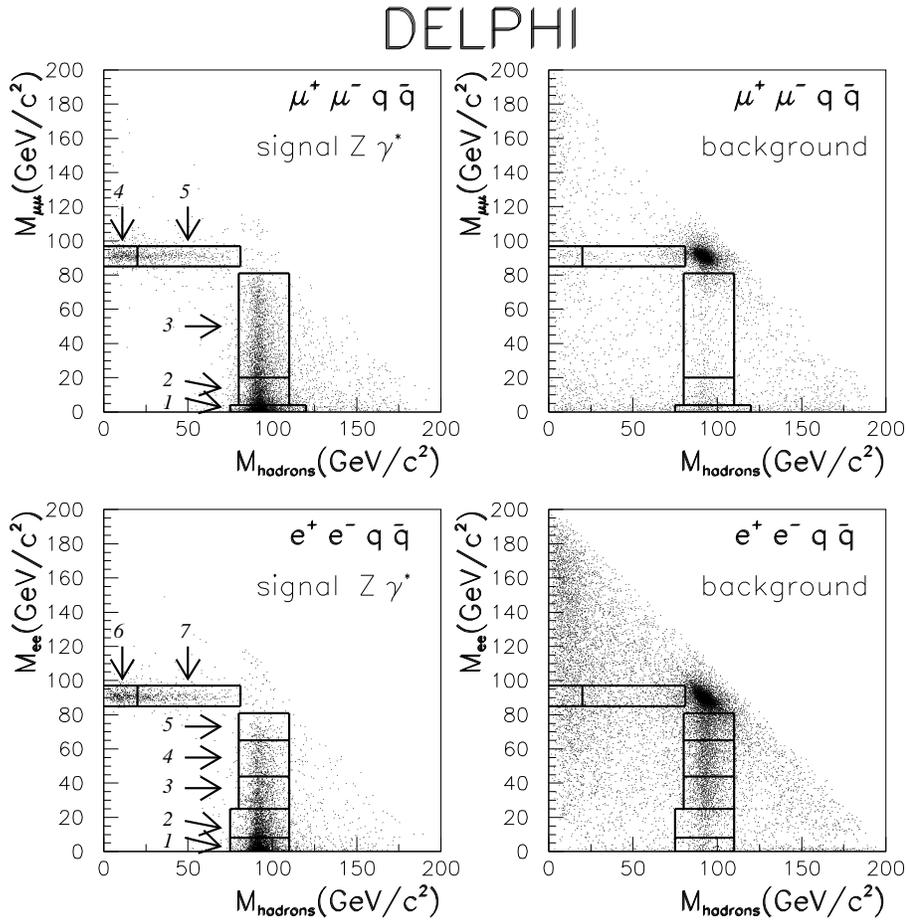}}
\caption{
$l^+ l^- q \bar{q}$ final state: Bidimensional distributions in the plane of the
di-fermion masses predicted by the Standard Model for signal $Z \gamma^*$
(left-hand  plots) and background (right-hand plots) for the two channels
studied, averaged over all energy points. The bins used for the fit are also shown.}
   \label{fig:llqq2}
  \end{center}
\end{figure}

The bidimensional distributions in the plane of the masses of the two  fermion pairs predicted by the Standard Model are shown in figure~\ref{fig:llqq2} for the two channels studied, separately for  $Z \gamma^*$ and background.  The
presence of non-resonant contributions, particularly of the type $Zee$ and $\gamma^* ee$, is clearly visible in the $e^+ e^- q \bar q$ case.  The distributions were binned as shown graphically in figure~\ref{fig:llqq2}, using a small number of irregularly sized bins. This allowed  the regions where most of the background is concentrated to be avoided, except for the $Zee$ contribution, while keeping as much signal as possible. Bin sizes were chosen in order to have an approximately equiprobable distribution of simulated events, with a finer binning in  $e^+ e^- q \bar q$  so as to follow better the more complicated structure of the background distribution. The observed and  predicted numbers of events selected by this procedure at each energy point  are reported in  table~\ref{tab:llqqres}.

On combining the data from all energy points, the efficiency and purity of the selected $\mu^+ \mu^- q \bar q$ sample were estimated from the simulation to be 42.0\% and 84.7\%, respectively. The background is composed of  $\mu^+ \mu^- q \bar q$ events outside the signal definition and of contributions from other final states. Interference effects in the  $\mu^+ \mu^- q \bar q$ channel in the region considered are negligible, as they account for less than 0.1\% of the $Z\gamma^*$ cross-section. The predicted composition of the background is shown in table~\ref{tab:llqqbck}.

In the $e^+ e^- q \bar q$ channel, the purity of the selected sample is estimated to be only 49.2\%, mostly because of the unavoidable $Zee$
background, while the efficiency was evaluated to be 24.3\%. Interference effects between $Z\gamma^*$ and other four-fermion processes were estimated to account for -15\% and are thus not negligible: they are mostly concentrated in the region of $Z\gamma^*$ - $Zee$ overlap.
The predicted composition of the background is shown in table~\ref{tab:llqqbck}.

\begin{figure}[htb]
  \begin{center}
   \mbox{\epsfysize=12cm\epsffile{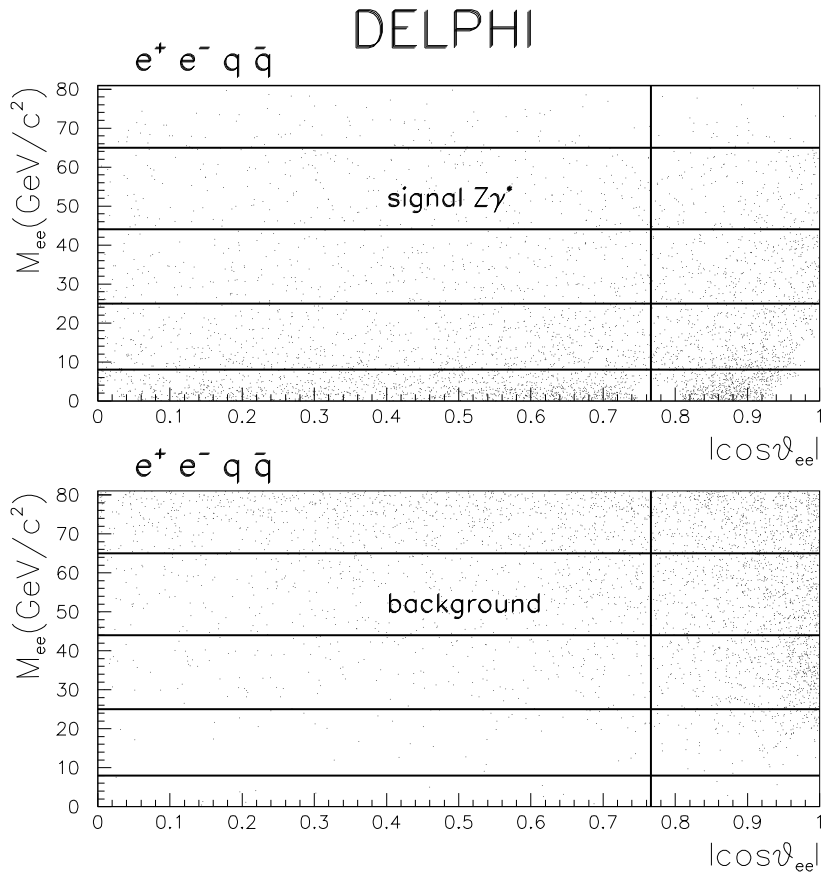}}
\caption{
Distributions of the mass of the electron-positron pair in $e^+ e^- q \bar q$ events for values of $M_{ee}$ less than 80~GeV/$c^2$ versus the polar direction of the pair  for signal $Z \gamma^*$ (upper plot) and background (lower plot) when the mass of the hadronic system is compatible with the $Z$ mass. The plot shows the predictions of the Standard
Model, averaged over all energy points. The binning adopted for these events follows that in figure~\ref{fig:llqq2} with an additional division into barrel and endcap regions, described in detail in the text.
}
   \label{fig:llqq3}
  \end{center}
\end{figure}

In order to disentangle the $Z \gamma^*$ from the $Zee$ contribution more
effectively, the distribution of the polar angle of the direction of the $e^+e^-$ pair was studied
as a function of the reconstructed invariant mass $M_{ee}$
in the  range defined by the first five  bins in figure~\ref{fig:llqq2}.
Correlation plots are shown in figure~\ref{fig:llqq3} for signal and background:
the distributions are well separated because in the $Zee$ case, which dominates
the background, even $e^+e^-$ pairs of large invariant mass are emitted at low
polar angles, due to the $t$-channel nature of the production process. The
binning in figure~\ref{fig:llqq2} was therefore modified,  as shown in
figure~\ref{fig:llqq3}, by doubling each bin, depending on whether a) the
polar angle of the direction of the $e^+e^-$ pair was in the barrel region ($40^\circ <
\theta_{ee} < 140^\circ$) or in the endcap region ($\theta_{ee} < 40^\circ$ or 
$\theta_{ee} > 140^\circ$) for bins 1-5, and b) the polar angle of the 
direction of the hadronic system was in the barrel or in the endcap 
region for bins~6-7. A total of 14 bins was thus used for the $e^+ e^- q \bar q$ 
cross-section measurement. This procedure resulted in an 8\% reduction of the
statistical error compared to the case where only mass bins were used.

\begin{table}[hbt]
\begin{center}
\begin{tabular}{|c|c|c|c|c|c|c|c|c|}
\hline
E (GeV) & \multicolumn{4}{c|}{$\mu^+ \mu^- q \bar{q}$} & \multicolumn{4}
{c|}{ $e^+ e^- q \bar{q}$  } \\
& Data & Total MC &Signal & Background & Data & Total MC & Signal & Background
\\ \hline
182.7         & 8 & 3.4  & 2.9 & 0.5  &
    4 & 3.3 &1.8  & 1.5  \\
188.6         & 8 & 9.3  & 7.8  & 1.5 &
    10 & 9.7 &4.6  & 5.1    \\
191.6      & 0 & 2.1  & 1.9  & 0.2  &
    1 & 1.4 &0.7  & 0.7    \\
195.5       & 2 & 4.1  & 3.5 & 0.6  &
    7 & 4.1 &2.1 & 2.0   \\
199.5       & 4 & 4.4  & 3.7  & 0.7  &
    5 & 4.1 &2.0  & 2.1    \\
201.6       & 3 & 2.1  & 1.8  & 0.3  &
    6 & 2.1 &1.0  & 1.1    \\
205.0       & 4 & 3.9  & 3.3  & 0.6  &
    1 & 4.1 &2.0  & 2.1    \\
206.5       & 6 & 7.4  & 6.2  & 1.2  &
    5 & 7.4 &3.6  & 3.8    \\
\hline
Total       & 35 &36.7 & 31.1  & 5.6  &
    39 &36.2 &17.8 & 18.4    \\
\hline
\end{tabular}
\caption []{ 
Observed numbers of events in the $\mu^+ \mu^- q \bar{q}$  and $e^+ e^-  q
\bar{q} $ channels at each energy compared with the Standard Model predictions
for signal and  background. }
\label{tab:llqqres}
\end{center}
\end{table}

\begin{table}[hbt]
\begin{center}
\begin{tabular}{|c|c|c|}
\hline
Background source & $\mu^+ \mu^-  q \bar{q} $  & $e^+ e^-  q \bar{q} $ \\ \hline
$WW$  & 0.8 & 1.6  \\ 
$q \bar q (\gamma)$  & 0.1 & 1.8  \\
$\tau^+ \tau^- q \bar q$  & 2.5 & 2.6  \\
non-$Z\gamma^*~l^+ l^- q \bar q$  & 2.2 & 18.0   \\
Interference  & $<0.001$ & -5.6  \\
\hline
Total  & 5.6 &18.4 \\ 
\hline
\end{tabular}
\caption []{Composition of the background to $Z\gamma^*$ production in the $\mu^+ \mu^- q \bar{q} $   and $e^+ e^-  q \bar{q}$ final states predicted by the Standard Model. The entries show the expected numbers of events, summed over all energy points. The row labelled non-$Z\gamma^*~l^+ l^- q \bar q$  shows the four-fermion neutral current
contributions from processes leading to the same final state as the signal, but defined as background, as described in section~\ref{sec:INTRO}.}
\label{tab:llqqbck}
\end{center}
\end{table}

\begin{figure}[p]
\begin{center}
    \mbox{\hskip 5truecm\epsfysize=10cm\epsffile{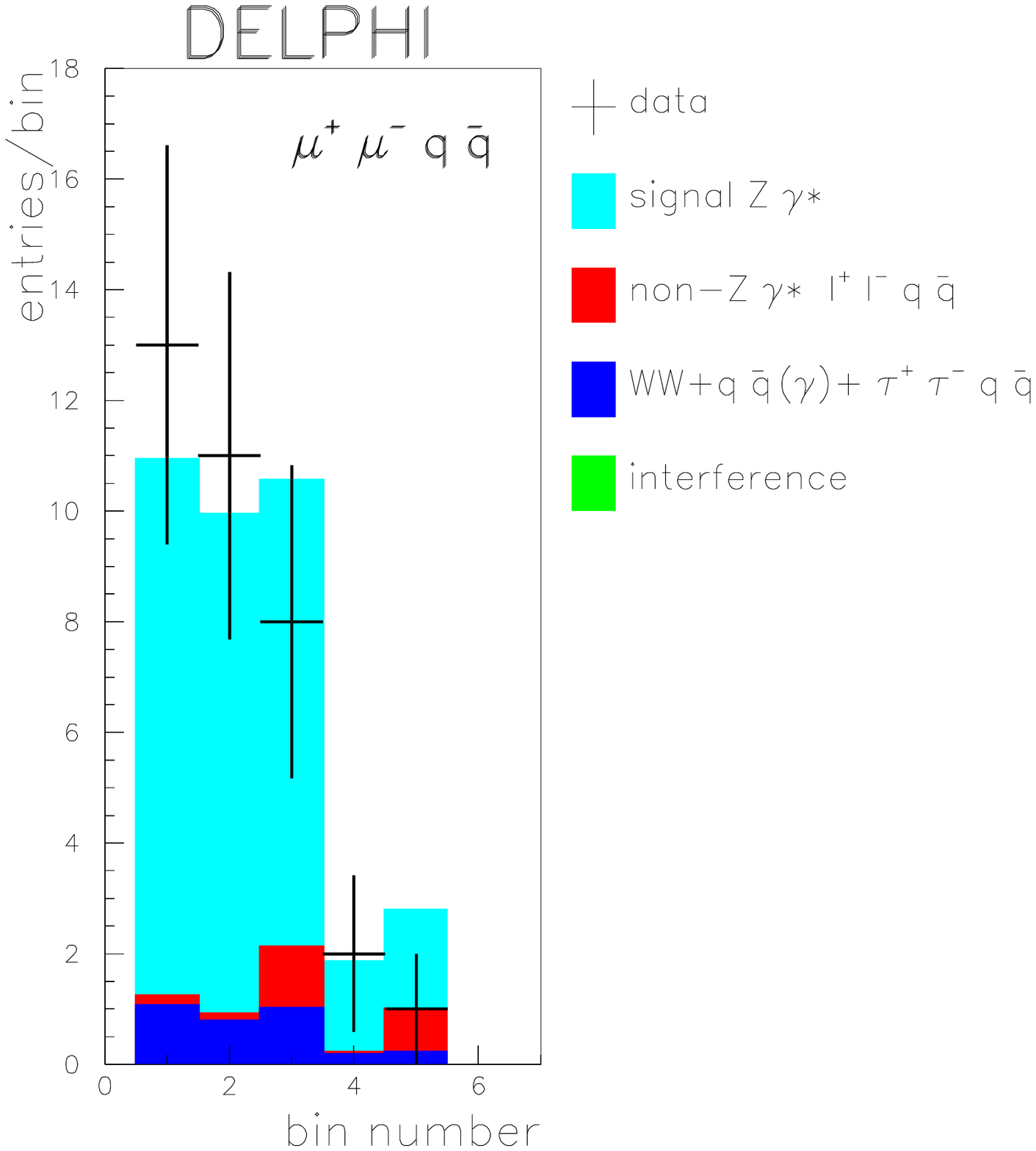 }}
    \mbox{\epsfysize=10cm\epsffile{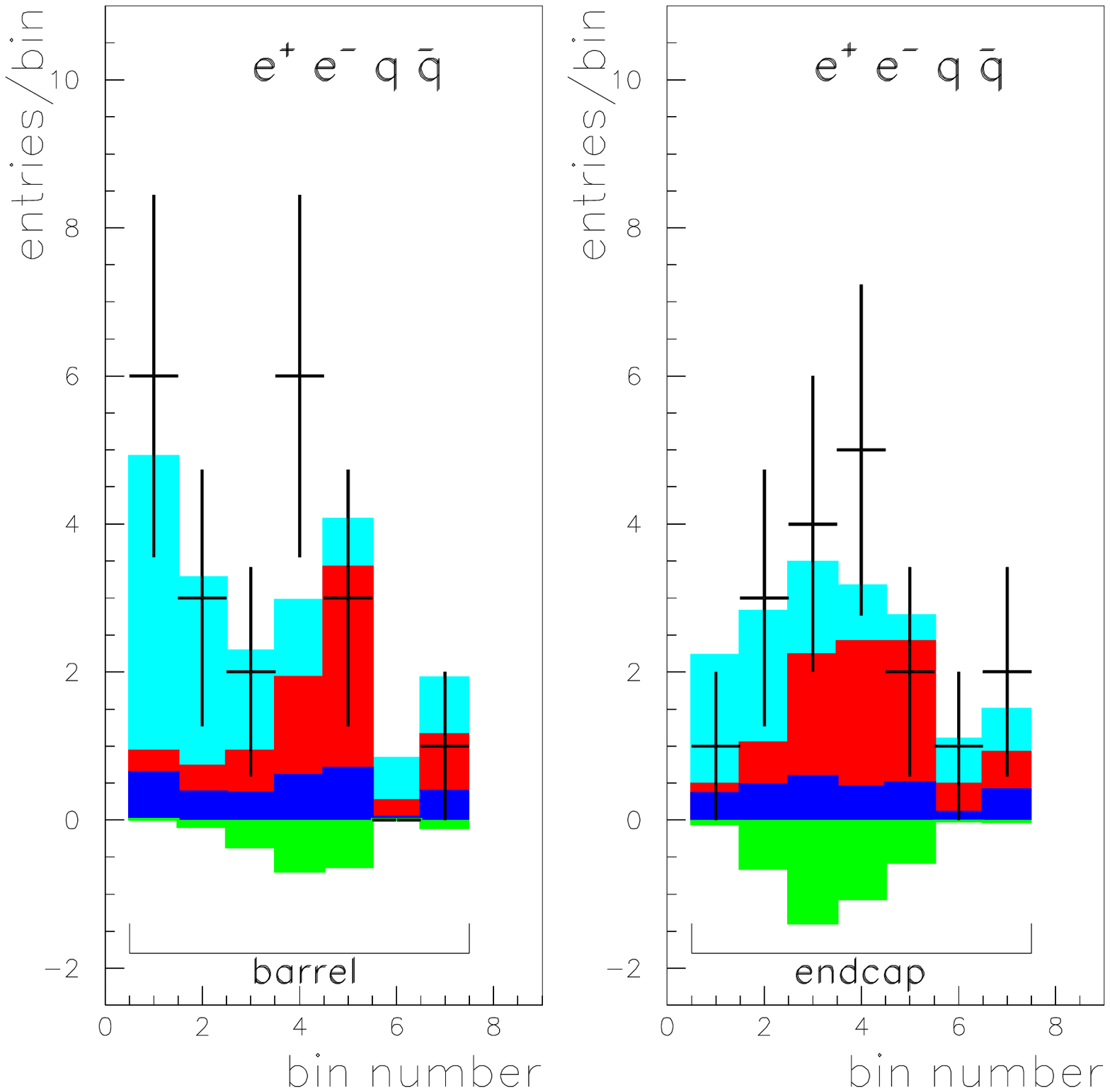 }}
\end{center}
\caption{
Results for $\mu^+ \mu^- q \bar q$ (upper plot) and  $e^+ e^- q \bar q$ (lower
plots). The points are data, summed over all energy points; the shaded 
histograms show the results of fits to the $Z\gamma^*$ component  
and the predicted Standard Model contributions from other sources. See figures~\ref{fig:llqq2}
and~\ref{fig:llqq3} and the main text for the bin definitions.}
\label{fig:llqq4}
\end{figure}

The value of the $Z\gamma^*$ cross-section at each energy point was extracted using a binned likelihood fit technique (see section~\ref{sec:RESULTS}) and the values were then  combined to get global results, separately for $\mu^+ \mu^- q \bar q$ and  $e^+ e^- q \bar q$.  Only the value of the $Z \gamma^*$ contribution was varied in the fit, while all non-$Z \gamma^*$ contributions, backgrounds and interference terms were fixed to the Standard Model expectations. 
Figure~\ref{fig:llqq4} compares the data in each bin used in the fits to the $\mu^+ \mu^- q \bar q$ and  $e^+ e^- q \bar q$ final states with the results of the fit, showing the contributions from the $Z\gamma^*$ signal, from the non-$Z\gamma^*$ component of each of the final states, from the interference terms, and from the other sources of background. 

These results were used to derive  the combined values of the $Z\gamma^*$
cross-section for the Matrix Element signal definition, as described in 
section~\ref{sec:RESULTS}. 

In the  $e^+ e^- q \bar q$ case, where  the presence of non-resonant diagrams is relevant, a two-parameter fit was also performed as a consistency check, leaving both the $Z\gamma^*$ and the non-$Z \gamma^*$ contributions free to vary, while fixing the remaining background sources and interference terms  to the Standard Model expectations. No significant change in the $Z\gamma^*$ cross-section result was observed, while the ratio $R_{non-Z \gamma^*}^{e^+ e^- q \bar  q}$ of the measured to the predicted cross-section of the non-$Z \gamma^*$ contribution to $e^+ e^- q \bar q$ was determined to be $R_{non-Z \gamma^*}^{e^+ e^- q \bar  q} = 1.15~^{+0.26}_{-0.23}$,  where the error is statistical only.

\subsection{Systematic errors \label{sec:LLQQSYST}}

Several sources of systematic error were investigated.

Uncertainties in lepton identification, signal efficiency and background levels were evaluated using a procedure similar to that in~\cite{ZZdelphi}, where the same final
states were studied.

Uncertainties in the lepton identification were estimated by comparing semileptonic
$WW$ events selected in data and simulation using the strong lepton
identification criteria. Uncertainties in signal efficiencies were evaluated by comparing the 
$P_t^{min}$ and $\chi^2$ distributions in data and simulation for all $l l q \bar q$ events
selected without mass cuts. Corresponding uncertainties in background levels
were evaluated by comparing samples of events selected in data and in
simulation, requiring both isolated tracks not to be identified as leptons,
while maintaining all the other criteria. Finally, uncertainties in the
background level in the $e^+ e^- q \bar{q}$ channel from fake electrons were
studied with $q \bar q (\gamma)$ events selected in data and in simulation with
purely kinematic criteria. These effects and the statistical uncertainty of
simulated data yielded a combined relative systematic error on the efficiency to select 
$\mu^+ \mu^- q \bar{q}$ and $e^+ e^- q \bar{q}$ events of $\pm 5.0\%$, and a
relative uncertainty in the background level of  $\pm 15\%$.\footnote{In both
cases determinations were limited in accuracy by the statistics of the available
samples, and should be interpreted as upper bounds.}

Systematic effects coming from the fitting procedure were 
investigated. Fit results were found to be stable within the expected 
statistical uncertainties against variations of bin sizes, the number
of bins and, for $e^+ e^- q \bar{q}$, the number of fitted parameters. 
No systematic error was thus attributed to this source.
 
Possible systematic effects arising from the treatment in the fit of
the $Z\gamma^*$  interference term with the other contributions (particularly the
non-$Z\gamma^*$ one) were taken into account (see section~\ref{sec:LLQQRESULTS}) for $e^+ e^- q \bar{q}$.  
Both the one-parameter and two-parameter fits were
repeated and the interference term was weighted with a factor proportional to
the product of the  $Z\gamma^*$ and non-$Z\gamma^*$ amplitudes. 
This changed the cross-section result by 2\%. Note, however, that this
procedure neglects a possible change in the phase between the two interfering
amplitudes with respect to that predicted in the Standard Model, and the 
procedure adopted therefore estimates the maximum possible effect that the
unknown phase could have. A systematic uncertainty of $\pm 2\%$ was thus ascribed from this source for
$e^+ e^- q \bar{q}$ events.

The systematic error coming from the uncertainty in the luminosity
measurement was evaluated to be $\pm 0.6\%$ both for  $e^+ e^- q \bar{q}$
and for $\mu^+ \mu^- q \bar{q}$.

The total estimated systematic errors on the measured  $Z\gamma^*$
cross-sections were  $\pm 5\%$ for $\mu^+ \mu^-  q \bar q$ and  $\pm 6\%$ for
$e^+ e^-  q \bar q$.

\section{Study of the $q\bar{q} \nu \bar{\nu}$ final state \label{sec:VVQQ}}

The  $q \bar q \nu \bar \nu$ channel is observed in a final state  topology of
hadronic matter and substantial missing energy. About half of the $Z \gamma^*$
cross-section in this channel  comes from the region of $q \bar q$ masses below
6~GeV/$c^2$.  Thus, final states often have the characteristic signature of
``monojets'', with the low invariant-mass hadronic system, which is the event
visible mass, arising from the $\gamma^*$ hadronisation and recoiling against a
highly energetic $\nu \bar \nu$ pair which escapes detection. 

Three analyses were performed and combined. The first analysis was intended to probe the low mass region of the hadronic system, so as to be efficient in the region of virtual photon mass, $M_{\gamma^*}$, below 6~GeV/$c^2$, where most of the cross-section is expected. It is denoted as the ``low mass analysis'' in the following. The second analysis exploited the large energy imbalance of $q \bar q \nu \bar \nu$ events, and retained some efficiency in the very low mass region of the hadronic system. It is denoted as the ``energy asymmetry analysis'' in the following. The third
analysis was intended to have good overall efficiency for high  $M_{\gamma^*}$ at the expense of having very small efficiency in the low  $M_{\gamma^*}$ region.  It is denoted as the ``high mass analysis'' in the following.

A common event preselection was defined for the three analyses, aimed mainly  at
reducing the backgrounds from two-photon and Bhabha events. The energy
measured in the electromagnetic calorimeters was required to be less than 60~GeV
in total and less than 10~GeV at polar angles below $15^\circ$ and above
$165^\circ$. Events with identified electrons at polar angles below $15^\circ$
and  above $165^\circ$ were excluded; the visible energy of the event was
required to exceed 15\% of the centre-of-mass energy; the polar angle of the
direction of the event missing momentum was required to be in the range
$15^\circ < \theta_{miss} < 165^\circ$; and at least two charged particles with
momentum greater than  200~MeV/$c$ were required.

An extensive use of veto counters was implemented in all three analyses: events with hits in the photon veto counters far from energy deposits in calorimeters or reconstructed tracks were rejected. The details of the algorithms adopted are given in the following sections.

In order to increase the available statistics, no explicit lower cut on the reconstructed mass of the hadronic system was applied. 

The numerical values of the cuts applied to kinematic variables  in the three analyses were chosen using an optimisation procedure described in section~\ref{VVQQRESULTS} below.

\subsection{Low mass analysis \label{VVQQ1}}

Events with a visible mass $M_{vis} < 6$~GeV/$c^2$ and with visible energy
$E_{vis}$ larger than 20\% of the centre-of-mass energy  were selected. In
addition, in order to limit the background from leptonic decays of $W$s
($W \rightarrow e/\mu~\nu, ~W \rightarrow \tau \nu, ~\tau\rightarrow e/\mu~\nu$),  
it was required that no identified muon be present, while at most one electron
was allowed in the event and its energy was required to be less than 30~GeV. 
Furthermore, events with the polar angle of the direction of the missing momentum in the range $38^\circ$ 
to $42^\circ$  (which is insufficiently covered by calorimeters, see section~\ref{sec:DETEC})  were rejected. 
The event was then  split into two hemispheres by the plane perpendicular to the thrust axis: events were rejected if there were hits in the photon veto counters in the hemisphere containing the direction 
of the missing momentum, while events with hits in the veto counters in the opposite
hemisphere were accepted only if their angular separation from the closest
charged-particle track or calorimetric energy deposit was less than  $20^\circ$.
  
When used alone, this analysis selected 10 events in data and 6.7 in the
simulation, of which 4.3 were signal and 2.4 were background.

\subsection{Energy asymmetry analysis \label{VVQQ2}}

In this analysis events were required to show a marked imbalance in the spatial distribution of the detected reaction products. Only events with total visible energy exceeding 20\% of the centre-of-mass energy were accepted. Then two
hemispheres were defined by a plane perpendicular to the direction of the thrust axis, and the total energy in each hemisphere was estimated from the curvature of charged-particle tracks and from calorimetric measurements. It was required that the energy in one of the two hemispheres account for at least 99\% of the total energy in the event. This was the main topological selection of the analysis and provided an implicit upper cut-off on the total visible mass of events.

Signals from photon veto counters were used to discard events with possible loss of
energetic photons in the insensitive regions of the electromagnetic calorimetry by adopting
the same algorithm as in the low mass analysis (see section~\ref{VVQQ1}).
In order to limit further the background from processes which have most of the
cross-section in the forward region (mainly Bhabha and two-photon events), the
cut on the polar angle of the direction of the missing momentum was tightened and
required to lie in the range $25^\circ < \theta < 155^\circ$.    

At this level, the background was completely dominated by the $WW$ and $W e \nu$
processes. In order to reject  leptonic decays of $W$s, events with
identified muons were discarded, while events with at most one electron were
kept if the energy of the electron did not exceed 25 GeV and the electron was
not isolated, i.e. its angle with respect to the closest charged-particle track was not
larger than $10^\circ$.   

Additional selections were implemented in order to suppress further the $WW$ 
and $W e \nu$ backgrounds. Part of this background arises from hadronic decays
of one $W$, accompanied by undetected leptons coming from the decay of the 
other $W$ or lost in the beam pipe (especially in the case of $W e \nu$). Such events
usually show larger visible mass than is expected from signal events, due to the
sizeable mass of the $W$ boson. A selection on the event visible mass was thus
imposed, requiring $M_{vis} < 45$~GeV/$c^2$. Another important fraction of
the remaining background comes from $WW$ events with both $W$s decaying to
$\tau$s, $W \rightarrow \tau \nu_{\tau}$, with the visible decay products
boosted into the same hemisphere. The signature of these events is that a few particles carry most
of the visible energy and have  visible mass above a few GeV/$c^2$. Two more
selections were imposed to reject such a source of background. Events with
visible mass above 5 GeV/$c^2$ and with more than 90\% of the visible energy
carried by the two most energetic particles were discarded. The remaining events
were forced  into two jets with the LUCLUS algorithm~\cite{luclus}.  Events with
total particle multiplicity below 11 and an angle between the two jets above
$30^\circ$ were rejected. 

When used alone, this analysis selected 25 events in data and 29.5 in the
simulation, of which 17.3 were signal and 12.2 were background. Half of the
background was contributed by the $WW$ and $W e \nu$ processes.

\subsection{High mass analysis \label{VVQQ3}}

In this analysis a cut on the multiplicity of charged-particle tracks was applied,
requiring it to be larger than 4. This implied that the efficiency of the
analysis dropped essentially to zero for $q \bar q$ masses below 2 GeV/$c^2$. 
The main topological selections were applied at jet level.  Jets were reconstructed using 
the LUCLUS algorithm and the events were forced into a two-jet configuration. An
upper cut on the opening angle of the two jets was set at $78^\circ$. The
parameter $d_2^{join}$ was defined to be the value for which the event passes
from a two-jet to a single jet configuration: only events with $d_2^{join} <
30$~GeV/$c$ were retained. The acoplanarity (defined as the complement of the
angle between the jets projected on the plane perpendicular to the beams) was
required to be larger than $90^\circ$.  Then the event was split into two
hemispheres about a plane perpendicular to the thrust axis and the energy
asymmetry, evaluated as in section \ref{VVQQ2}, was required to be larger than
95\%. Events with missing mass less than 80~GeV/$c^2$ were rejected. 

A further selection was imposed on the energy of the
visible system, $E_{vis}$, rejecting events with  $E_{vis}
>$~80~GeV. In the absence of initial- and final-state radiation, the energy and
the mass of the $q \bar q$ system in the $Z \gamma^*$ process are related in 
the following way:  
$$ E_{q \bar q} = \frac{s-M^2_Z+M^2_{q \bar q}}{2 \sqrt{s}}~. $$ 

\noindent The quantity $E_{kin} = \frac{s-M^2_Z+M^2_{vis}}{2 \sqrt{s}}$  was
defined, using the visible mass of the event. It was then required that the
difference between $E_{kin}$ and the visible energy of the event $E_{vis}$
did not exceed 45 GeV. This cut, and the cut on $E_{vis}$ described above, were
effective in suppressing the $WW$ and $q \bar q (\gamma)$ backgrounds.  

Events with hits in the photon veto counters were accepted if the angular distance between these hits and the direction of the closest jet was less than $30^\circ$; otherwise they were rejected. 

When used alone, this analysis selected 21 events in data and 20.7 in the
simulation, of which 13.4 were signal and 7.3 were background. Most of the
background is due to $WW$ and $W e \nu$ events.

\subsection{Results \label{VVQQRESULTS}}
The three analyses were combined on an event-by-event basis, by selecting events which passed any of the three selections. Numerical values of the cuts were optimised in a two-stage procedure. First, for each analysis separately, all the cuts relevant to that analysis were varied such that the product of efficiency and purity of the selected sample was  maximised. Then the most important cuts in each analysis were allowed to vary simultaneously, keeping the other cuts at the values obtained in the first stage, and the product of the efficiency and purity of the sample selected by any of the three analyses was maximised. (The values listed in sections~\ref{VVQQ1}, \ref{VVQQ2}~and  \ref{VVQQ3} are the result of this last optimisation procedure).  In total, 42 events were found in data and 41.3 in the simulation (with a total overlap between the three selections of about 30\%); of the simulated sample, 23.4 events were signal and 17.9 were background. The most abundant source of background was predicted to come from $W e \nu$ events,  which accounted for 7.9 events, mainly in the channel $q \bar q e \nu$ and partially in $\tau \nu e \nu$. On-shell $WW$ processes contributed about 4 events to the background, with 2.9 of them containing at least one $W$ decaying to $\tau \nu$. The remaining main sources of background were $q \bar q$ (about 2 events), $\tau \tau$ (about 2 events) and other four-fermion neutral current processes (1.1 events). Table~\ref{tab:vvqqres} shows the numbers of signal and background events predicted by the Standard Model and the observed numbers of events in the $q \bar{q} \nu \bar \nu$ channel at the various centre-of-mass energies. 

\begin{table}[hbt]
\begin{center}
\begin{tabular}{|c|c|c|c|c|}
\hline
E (GeV) & Data & Total MC &Signal & Background \\ \hline 
182.7  & 3 & 3.5  & 2.3 & 1.2 \\ 
188.6  & 9 & 10.1 & 6.0 & 4.1 \\
191.6  & 1 & 1.3  & 0.9 & 0.4 \\
195.5  & 7 & 4.4  & 2.9 & 1.5  \\
199.5  & 6 & 5.2  & 2.9 & 2.3  \\
201.6  & 2 & 2.5  & 1.3 & 1.2  \\
205.0  & 9 & 4.9  & 2.6 & 2.3  \\
206.5  & 5 & 9.4  & 4.5 & 4.9  \\
\hline
Total  & 42 &41.3 & 23.4  & 17.9 \\ 
\hline
\end{tabular}
\caption []{ 
Observed numbers of events in the $q \bar q \nu \bar \nu$ channel at each energy
compared with the Standard Model predictions for signal and  background. }
\label{tab:vvqqres}
\end{center}
\end{table}

\begin{figure}[h]
  \begin{center}
   \mbox{\epsfysize=13cm\epsffile{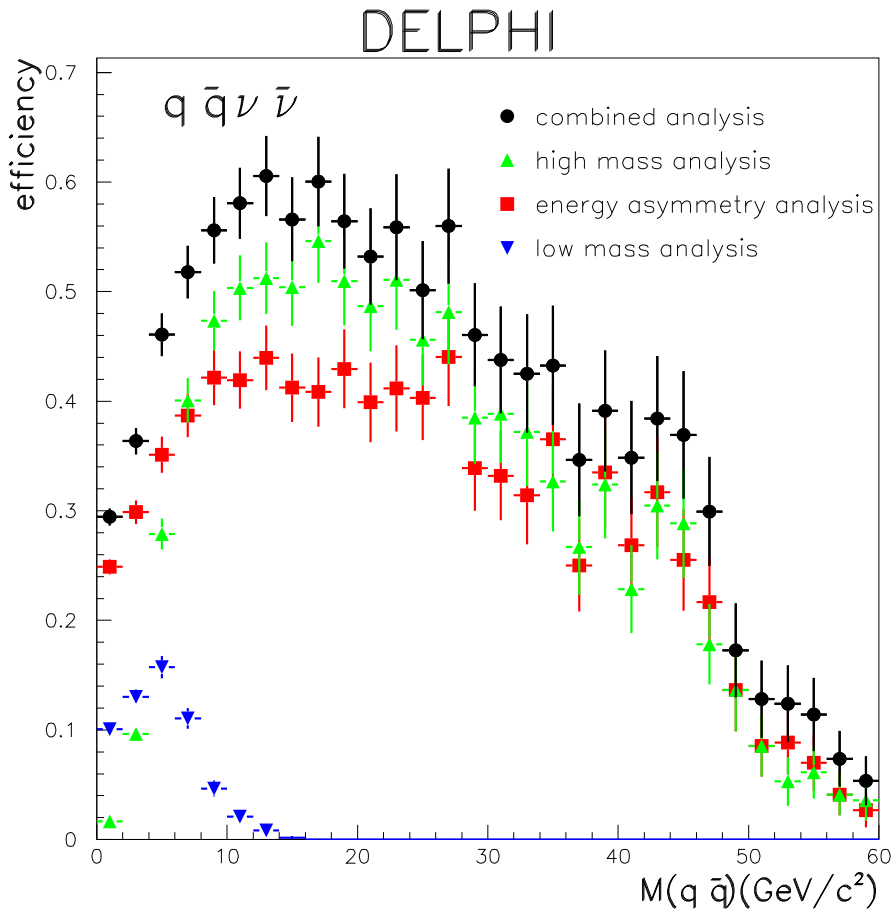}}
\caption{Selection efficiency of the  $q \bar q \nu \bar\nu$ analyses, averaged over all energy points, as a function of the generated $M(q \bar q)$ mass. The efficiency is shown for each of the three analyses (see text) separately, and for the combined analysis.}  
   \label{fig:vvqq2}
  \end{center}
\end{figure}

\begin{figure}[h]
  \begin{center}
   \mbox{\epsfysize=13cm\epsffile{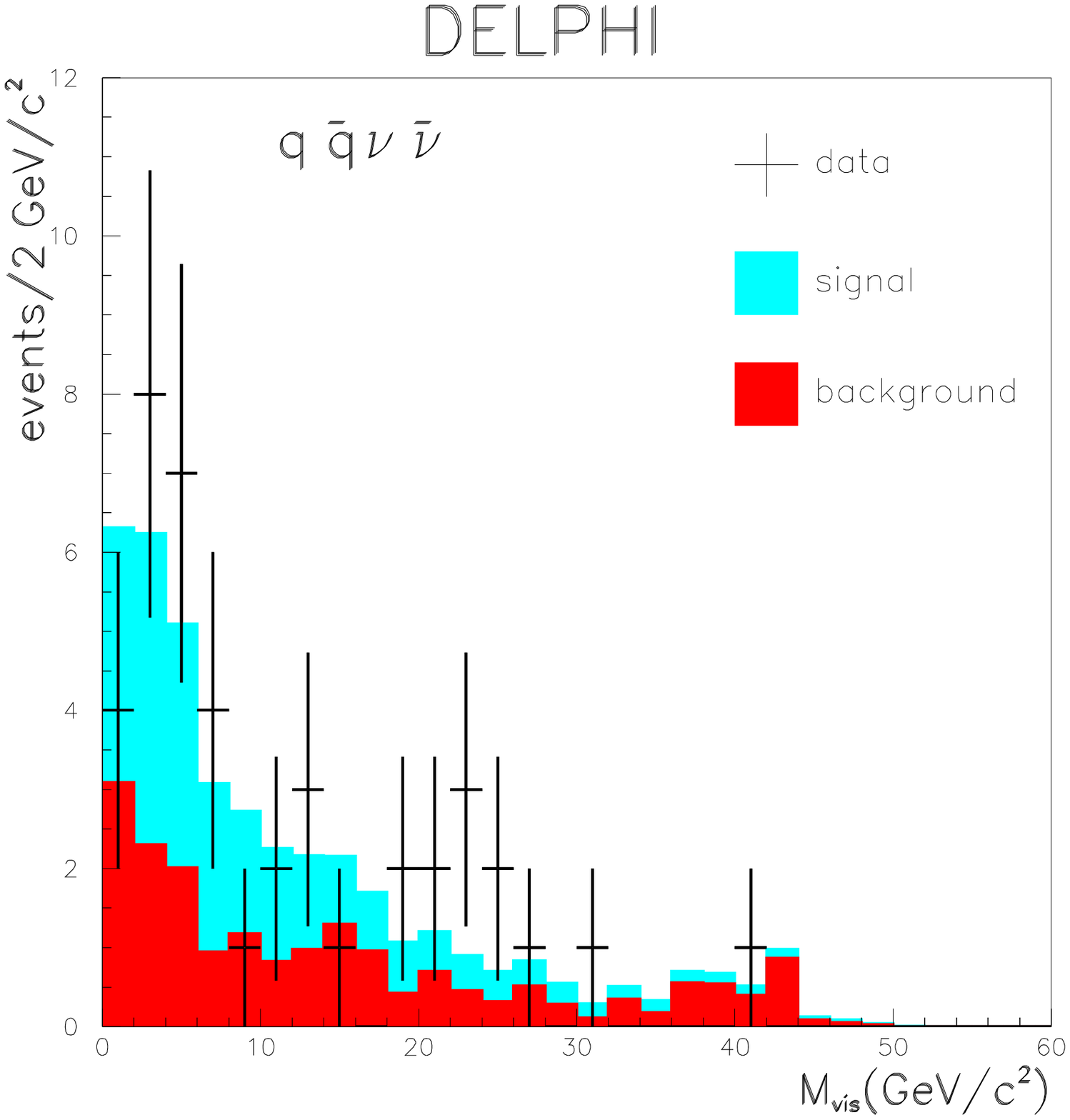}}
\caption{
Distribution of the visible invariant mass of the hadronic system in the $q
\bar{q} \nu \bar \nu$ selection, compared with the Standard Model predictions
for signal and background. The points are the data, summed over all energy 
points, the light (light blue) histogram shows the predicted signal 
contribution, and the dark (red) histogram
shows the predicted background.}
   \label{fig:vvqq1}
  \end{center}
\end{figure}

The differential efficiencies of the three analyses as a function of the generated mass, $M(q \bar q$),  estimated from the simulation, are shown in figure~\ref{fig:vvqq2}, together with the efficiency for the combined selection. The overall selection efficiency, averaged over all masses, was estimated to be 38.8\%. The distribution of the reconstructed  visible mass,  $M_{vis}$, for the 42 data events is shown in figure~\ref{fig:vvqq1}, which also shows the distributions for the simulated signal and background events. Good agreement is observed with the Standard Model expectations.

The value of the $Z \gamma^*$ cross-section at each energy point was extracted
using a counting technique and the values were  then combined to get a global
result. All non-$Z \gamma^*$ contributions, backgrounds and interference terms
were fixed to the Standard Model expectations. The result was used to derive a
combined value for the $Z \gamma^*$ cross-section in the Matrix Element signal 
definition, as described in section~\ref{sec:RESULTS}.  

\subsection{Systematic errors \label{VVQQSYST}}

Various sources of systematic error were considered. 

The predicted background cross-sections were varied according to the following
values: $WW:\pm 2\%$,  $q \bar q:\pm 5\%$,  $W e \nu:\pm 5\%$,  $\tau \tau:\pm
5\%$, four-fermion neutral current processes: $\pm 5\%$. The combined effect on
the cross-section measurement was estimated to amount to $\pm 2\%$, with the
main contribution coming from the uncertainty on the $W e \nu$ cross-section. 

Uncertainties on the signal efficiency coming from the Monte Carlo generator
were studied by comparing different generator models. In particular, a sample of
$q \bar q \nu \bar \nu$ was generated with the EXCALIBUR~\cite{excal}
four-fermion generator for masses of the hadronic system $M(q \bar q) > 10$~GeV/$c^2$. 
For generated masses below 10~GeV/$c^2$ the hadronisation model in EXCALIBUR is not as reliable as that in WPHACT and systematic effects from that region were evaluated separately (see below).
The full analysis was applied to the EXCALIBUR sample and a difference of 3\% in
the signal efficiency was obtained.  A systematic  uncertainty of $\pm 3\%$ was
thus conservatively ascribed to this source.    

Systematic uncertainties due to the description of the hadronisation mechanism in the $q \bar q$ system were taken into account. It was assumed that these effects can be relevant for masses  $M(q \bar q) < 10$~GeV/$c^2$ 
(see above), affecting the analysis mainly through corresponding uncertainties in charged-particle multiplicity distributions. These effects were not expected to be large because two of the three analyses (the low mass
analysis and the energy asymmetry analysis) adopted a very low cut on the charged-particle multiplicity. The study of these effects was split into two parts, corresponding to the resonance and the continuum contributions respectively (see section~\ref{sec:DETEC}). In the simulated sample the dominant resonances were identified, their corresponding detection efficiencies computed, and their contributions varied by amounts derived from the uncertainties in
their known measured cross-sections: 1\% for $\rho$ and $\phi$ production, 10\% for resonances decaying to final states with 3 or 4 charged particles, and  30\% for resonances decaying to states with 5 or 6 charged particles.
The effect on the estimated cross-section was found to be negligible; this is not surprising as $\rho$~production, for which the cross-section is accurately determined, accounts for about 80\% of the cross-section below 2 GeV/$c^2$. 
As a second step, the contribution of the resonances was subtracted from the hadronic mass distribution, the 
charged-particle multiplicity distribution of the remaining sample studied and the analysis efficiency evaluated
as a function of the number of charged particles. The effect of a possible error in the determination of the 
charged-particle multiplicity distribution was then estimated by stretching the observed distribution by $+20\%$, rebinning, and applying the efficiency curve to the new distribution. The procedure was repeated, compressing the distribution by 20\%. The range of cross-sections obtained from the stretched and compressed distributions was taken as an estimate of the systematic error. The effect on the $Z \gamma^*$ cross-section amounted to $\pm 4\%$. 
  
\begin{figure}[h]
  \begin{center}
   \mbox{\epsfysize=13cm\epsffile{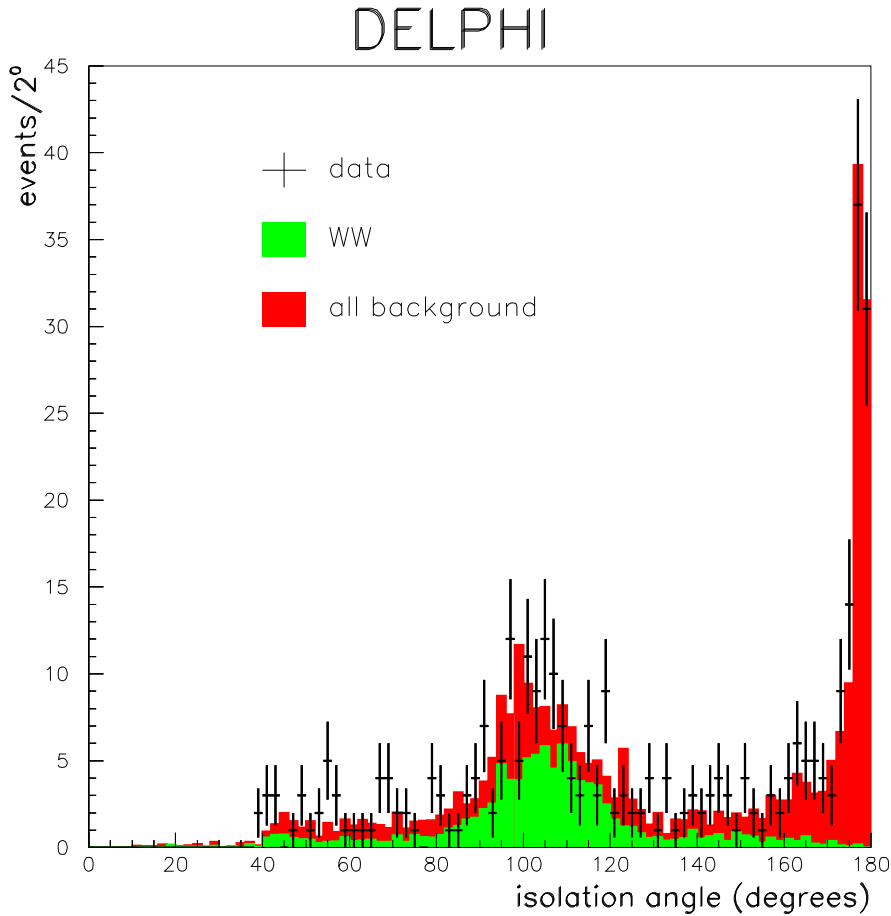}}
\caption{Isolation angle of selected electrons, muons and single charged-particle tracks in the sample selected to mimic the background to the selected $q \bar q \nu \bar \nu$ sample. The points are the data, summed over all energy 
points, the light (green) histogram is the predicted $WW$ contribution, and the dark (red) histogram is the rest of the background.} 
   \label{fig:vvqq3}
  \end{center}
\end{figure}

Another source of systematic uncertainty considered was the reliability of the
simulation in correctly estimating the amount of background.  As explained in
section~\ref{VVQQRESULTS}, the main backgrounds are $WW$ events, with one or
both $W$s decaying to $\tau$, and $W e \nu$ events, with the on-shell $W$
decaying hadronically or to $\tau \nu$.   These events share the common feature
of having the decay products of one $W$ detected on one side, and missing
energy on the other side.  The missing energy is due to the low angle electron,
typically lost in the beam pipe in the $We\nu$ case, or to an undetected decay
lepton or charged-particle track in the $WW$ case. Furthermore, in both topologies,
additional missing energy is carried by the escaping neutrino. In order to
evaluate the reliability of the simulation in estimating the efficiency to
detect backgrounds in such a topology,   events with features similar to 
those of the background in the $q \bar q \nu \bar \nu$ analysis were studied.  In particular, $WW$ events with one
$W$ decaying to a detected lepton (electron or muon) or to an isolated 
charged-particle track, which was then artificially removed from the event,  can mimic most of
the $WW$ and $We\nu$ background, with the second $W$ playing the role of the
hadronic signal. Therefore events with an isolated electron, muon or other
charged-particle track were selected. Identified leptons or other charged-particle tracks
were initially required to have momentum    larger than 10~GeV/$c$ and an angle
with respect to the closest charged-particle track larger than $10^\circ$. The selected
candidate track was then excluded from the event and the selections in
sections~\ref{VVQQ1}, \ref{VVQQ2} and  \ref{VVQQ3}  applied to the remaining
system. At the end of the procedure, 142 events with an isolated muon were found
in data and 135.7 in the simulation, 110 events with an isolated electron were
found in data and 115.1 in the simulation, and 79 events with a single isolated
charged-particle track were found in data and 72.4 in the simulation. The distribution
of the isolation angle of the selected lepton or single track after all the cuts
is shown  in figure~\ref{fig:vvqq3}. Good agreement between data and simulation
is observed. The dominant contributions to the events selected in this way
come from semileptonic $WW$ production, $\tau \tau$ events and, to a lesser
extent, Bhabha events, and can thus be used  to emulate the background to the 
$q \bar q \nu \bar \nu$ signal:  when the isolated lepton or other charged-particle track is excluded from the sample,  the remaining system is strongly asymmetric in the angular distribution of the visible momentum, and of the same topology as the background expected in the $q \bar q \nu \bar \nu$ sample. (The estimated contribution from the $q \bar q \nu \bar \nu$ signal to this sample is totally negligible).

The agreement between data and simulation in figure~\ref{fig:vvqq3} was subjected to more detailed checks, for example by selecting the region of the distribution in the isolation angle of the single charged-particle track  which enriches the sample in $WW$ events:  two-fermion events preferentially  populate the region of large isolation angle, being almost back to back, and their contribution can be greatly reduced with a cut at around 130$^\circ$. Similarly, other checks were made for different  visible-mass and track-multiplicity regions; in all the cases the agreement between data and simulation was good within the errors. The statistical error in the total of 331 events selected by this procedure  was thus taken as an estimate of the systematic uncertainty due to the background evaluation from the simulation; this gave a contribution of~$\pm 3\%$ on the $q \bar q \nu \bar \nu$ cross-section measurement.  

Systematic uncertainties from the trigger efficiency were investigated and found to be negligible: the triggering efficiency for a single charged-particle track with transverse momentum $p_T > 3$~GeV/$c$ is already very well determined~\cite{trig}, while in the present analyses a charged-particle track multiplicity of at least 2 was required, with transverse momenta of selected events in general well in excess of 3~GeV/$c$.

The systematic uncertainty coming from the luminosity measurement was
estimated to give an error of~$\pm 0.6\%$ on the cross-section measurement. 

The statistical error from the limited simulated sample gave an uncertainty of~$\pm 5\%$.

Finally, the stability of the result as a function of the applied experimental cuts was
checked by varying the selections of the three analyses, first separately and
then at the same time. The procedure set up to maximise the product of the 
efficiency and purity of the simulated sample (see section~\ref{VVQQRESULTS})
was used to vary all the relevant cuts of sections~\ref{VVQQ1}, \ref{VVQQ2} and 
\ref{VVQQ3} within reasonable values; selections were accepted if the predicted value for 
the selected sample differed by less than the  statistical
error of the optimum value obtained from the simulated sample used in the
analysis.  For each such selection, the background level and number of 
events in data were estimated and a value for the $q \bar q \nu \bar \nu$ 
cross-section was measured. The root mean square of the distribution of the cross-sections evaluated 
in this way was estimated to amount to 3\%. As this number is compatible with the statistical fluctuations intrinsic to this procedure, no systematic error was added.

The total estimated systematic error on the $Z\gamma^*$ $q \bar q \nu \bar
\nu$ cross-section measurement was thus estimated to be $\pm 8\%$.  

\section{ Study of the \llll\ final state  \label{sec:4LEPT}}

The Feynman diagrams of figure~\ref{fig:zgconv} give rise to six possible final
states with four charged leptons: \mmmm, \eeee, \tttt, \eemm, \eett\ and \mmtt.
These final states have a rather clean experimental signature, but do not
contribute significantly to the total four-fermion production cross-section due
to the low branching fraction of  $Z/\gamma^* \rightarrow l^+ l^-$. 

The selection of events in the \llll\ final state was restricted to topologies with four
 well reconstructed charged particles with momenta larger than 2~GeV/$c$ 
(henceforth called lepton candidates). Events with two additional well-measured 
charged particles with opposite charges were allowed, provided that the pair was 
compatible  with a photon conversion, or that the momentum of both particles was 
less than 2~GeV/$c$. Five additional charged particles were allowed in the event if their tracks did
not point to the vertex; such tracks were not considered in the following steps
of the analysis. The previous selections implied that for \eett, \mmtt and
\tttt\ events only one-prong $\tau$ decays were considered. The sum of the
charges of the lepton candidates had to equal zero and the angle between the
directions of any two of them had to be larger than 5$^\circ$.

The four lepton candidates were required to fulfil the following additional 
selection criteria: the momenta of at least three of them had to exceed 
6~GeV/$c$, their total energy had to be greater than 0.25$\sqrt{s}$  (to reject
background from two-photon interactions), and the length of at least 
three of the candidates' tracks was required to be greater than 50~cm. Beam-gas 
and $\tau^+\tau^-\gamma$ events were rejected by requiring that the four lepton candidates were not 
all in the same hemisphere with respect to the beam direction.   For data taken
during 2000, in the period when one sector of the TPC was not working, a
slightly  more relaxed criterion for track selection was applied if the track
traversed that  sector. 

Selected events in the data were compared with simulated signal and background
samples generated at the eight centre-of-mass  energies.  The expected numbers
of events for signal and background, together with the numbers of events found
in data, are shown in table~\ref{tab:4l1} both for the full sample of \llll\ events, selected 
as described above, and for the $Z\gamma^*$ sample defined in section~\ref{sec:llllres}.  
The  overall \llll\ selection efficiency is $\sim$15\%, increasing slightly with  $\sqrt{s}$ for the full sample, while 
for the $Z\gamma^*$ selection it ranges between 22\% and 30\%. The most 
important contribution to the non-\llll\ background comes from  $e^+e^- \rightarrow e^+e^- q \bar{q}$
events with low $q \bar{q}$ mass. The second most important contribution is due
to the $e^+e^- \rightarrow \tau^+\tau^-(\gamma)$ process. Good agreement was
found between the data and the predictions of the simulation after each selection was applied sequentially. 
For the $Z\gamma^*$ sample, the main background is due to \llll\ contributions 
from non-$Z\gamma^*$ processes.

\begin{table}[hbt]
\begin{center}
\begin{tabular}{|c|c|c|c|c|c|c|c|c|}
\hline
E (GeV) & \multicolumn{4}{c|}{$l^+l^-l^{'+}l^{'-}$ full sample} & \multicolumn{4}{c|}{$l^+l^-l^{'+}l^{'-}$ $Z\gamma^{*}$ sample} \\
& Data & Total MC &Signal & Background & Data & Total MC & Signal & Background
\\ \hline
182.7   &  3 &  3.9 &  3.4 & 0.4  & 1 & 1.5 & 0.5 & 1.0  \\
188.6   & 14 & 12.4 & 10.0 & 2.4  & 2 & 4.8 & 1.6 & 3.2  \\
191.6   &  1 &  1.8 &  1.6 & 0.2  & 1 & 0.6 & 0.2 & 0.4  \\
195.5   &  5 &  5.3 &  4.6 & 0.7  & 2 & 1.9 & 0.6 & 1.3  \\
199.5   &  8 &  6.0 &  5.1 & 0.8  & 2 & 2.0 & 0.7 & 1.3  \\
201.6   &  3 &  2.7 &  2.4 & 0.4  & 2 & 0.9 & 0.3 & 0.6  \\
205.0   &  7 &  5.3 &  4.8 & 0.5  & 4 & 1.8 & 0.7 & 1.2  \\
206.5   &  7 &  9.6 &  8.2 & 1.4  & 3 & 3.1 & 0.9 & 2.1  \\
\hline
Total   & 48 & 47.0 & 40.1 & 6.8  &17 &16.6 & 5.5 &11.1  \\
\hline
\end{tabular}
\caption []{
Observed numbers of events in the $l^+l^-l^{'+}l^{'-}$ channel for the full sample
and for the $Z\gamma^{*}$ sample at each energy, compared with the Standard
Model predictions for signal and background. In the case of the $Z\gamma^{*}$ sample, the background contributions are defined to include the non-$Z\gamma^*$ \llll\ contribution and the non-\llll\ contribution.}
\label{tab:4l1}
\end{center}
\end{table}

\subsection{Particle identification and final state classification
\label{sec:PAID}}

Events selected in the \llll\ final state were classified into one of the 
six final states according to the number of identified muons, electrons and 
pions. A constrained fit procedure was also used to complete the identification. 

Muon identification was performed by combining the standard DELPHI
identification package~\cite{delphi2} in the muon chambers with the energy
deposition  profile in the hadron calorimeter and the energy deposited in the 
electromagnetic calorimeter. 

Electron identification required that there be no signal in the muon chambers
and no energy deposited in the hadron calorimeter after the first  layer. The
energy in the electromagnetic calorimeter in a 2$\dgree$ cone surrounding the
candidate particle was required to be larger than  1~GeV. For electrons 
satisfying these criteria, the momentum of the charged particle was replaced by
the energy deposited in the electromagnetic calorimeter.

Pions were identified as tracks leaving an energy deposit in the electromagnetic
calorimeter compatible with a minimum ionizing signal, no hits in the muon
chambers and energy deposited in the layers of the hadron calorimeter
compatible with the profile of a hadron shower. 

The assignment of the final state proceeded as follows: 

\begin{itemize}

\item  If no $e^+e^-$ or $\mu^+\mu^-$ pair was identified, the four particles
were considered as $\tau$ decays and the final state to be \tttt;

\item If two pairs were identified as $e^+e^-$, $\mu^+\mu^-$ or  $\tau^+\tau^-$, 
the final state was considered to be fully identified;

\item If one $e^+e^-$ or  $\mu^+\mu^-$ pair was identified and the second pair
had two identified particles, different from one another, the event was
considered to be \eett\ or \mmtt, respectively. The second pair was also
designated as $\tau^+\tau^-$ if only one particle was identified and was
different from the identified pair, or if neither was identified;

\item If the event had 3 identified electrons or muons and one unidentified
particle, two hypotheses were considered: that the 4 particles were identical
or that the unidentified particle was one of a $\tau^+\tau^-$ pair. 

\end{itemize}

\begin{figure}[p]
\begin{center}
    \mbox{\epsfysize=17cm\epsffile{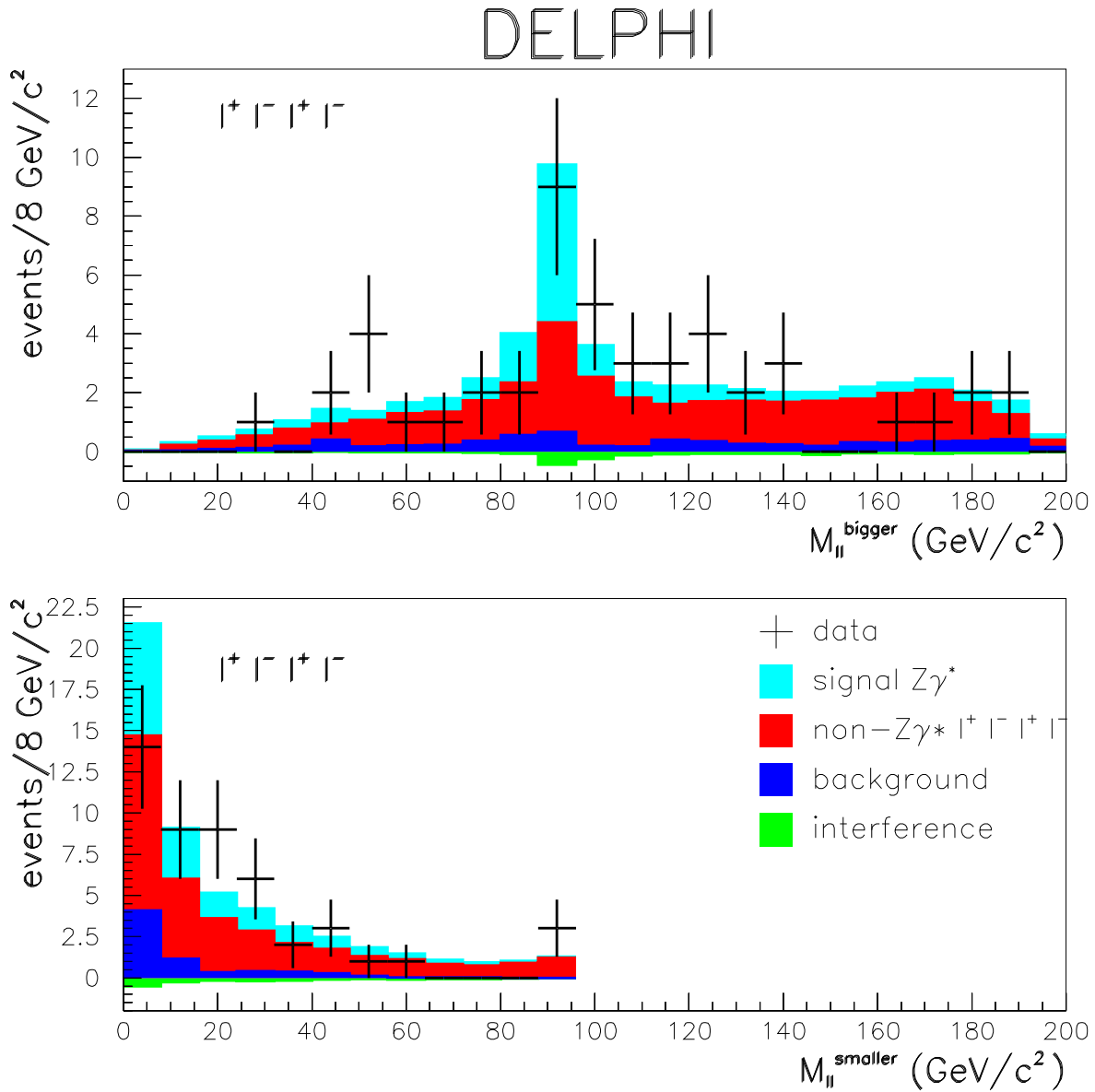}}

\end{center}
\caption{
Four-lepton channel:  Fitted masses of two selected lepton pairs: bigger mass
(top plot), smaller mass (bottom plot), compared with Standard Model
predictions. The points are the data, summed over all energy points, 
and the histograms represent the predicted
contributions to the selected event sample. In the legend, ``background"  means
the contribution from non-\llll\ final states.}
\label{fig:mass4l1}
\end{figure}

A constrained kinematic fit was then performed on the selected events, imposing four-momentum conservation. This implies a four-constraint fit in the case where both lepton pairs are either electrons or muons, and a two- or zero-constraint fit in the cases where, respectively, one or two tau pairs are assumed present, as the magnitude of the 
tau momentum was taken to be unknown. Where more than one kinematic hypothesis could be applied to the same event, the decision procedure and the final identification were based on the probability of the $\chi^2$ of the fit and the relative errors of the fitted masses. In the case of four identical particles, the combination for which a pair of leptons had reconstructed mass within 15 GeV/$c^2$ of the nominal $Z$ mass was chosen or, if this condition was not fulfilled, the combination with the largest fitted invariant mass of a pair of leptons was selected. If no acceptable hypothesis was found, further fits were tried where kinematically possible, assuming, in addition to the four leptons, the presence of an unobserved ISR photon in the beam pipe; again the best resulting fit was selected. Figure~\ref{fig:mass4l1} shows the distributions of the larger and smaller mass pairs for the full data sample, calculated for each event from the results of the chosen fit, and compares them  with the predictions of the Standard Model.

The efficiencies for assigning the correct final states to the selected events 
were estimated from the simulation. The results are summarised 
in table~\ref{tab:idmatrixall}, which shows the expected numbers of events from the 
full \llll\ sample which were identified in each of the possible final states, as 
well as the efficiency and purity.

\begin{table}[htbp]
\begin{center}
\footnotesize{
\begin{tabular}{|l|c|c|c|c|c|c|c|c|c|c|} 
\hline
 \ \  Identified   &\multicolumn{8}{c|}{Generated final state}&Data& Purity  \\
\cline{2-9}  
\ \ final state      &\eeee&\eemm&\eett&\mmmm&\mmtt&\tttt &Total&Bck& & (\%) \\ \hline
\ \eeee&6.4 & 0.0& 0.3& 0.0& 0.0& 0.0 &  6.7 & 1.4 &  7 &80\\ \hline
\ \eemm&0.0 &13.8& 0.3& 0.0& 0.1& 0.0 & 14.2 & 0.4 & 14 &95\\ \hline
\ \eett&3.2 & 1.5& 3.2& 0.0& 0.0& 0.1 &  8.0 & 3.0 & 16 &29\\ \hline
\ \mmmm&0.0 & 0.0& 0.0& 2.3& 0.1& 0.0 &  2.4 & 0.1 &  2 &92\\ \hline
\ \mmtt&0.0 & 4.0& 0.0& 0.4& 1.6& 0.0 &  6.0 & 0.4 &  7 &25\\ \hline
\ \tttt&0.4 & 0.5& 0.6& 0.0& 0.2& 0.2 &  1.9 & 1.5 &  2 & 6\\ \hline \hline
Efficiency (\%)&10 &11 &11 &13 & 12 & 9 &\multicolumn{4}{c|}{} \\
\hline
\end{tabular}
}
\caption{Upper six rows: Expected numbers of signal and background events and purity for each identified final state for the full \llll\ event sample, estimated from the simulation. The number of events found in the experimental data is also given for each final state.
Bottom row: Estimated  efficiency for selection and correct classification of each \llll\ state with respect to the total \llll\ content of the sample.}
\label{tab:idmatrixall}
\end{center}
\end{table}

\noindent In 2\% of the cases the events could not be classified in any of the six final states, as there was no complete {\it a priori} identification of all the particles in the event and the constrained fit failed. Due to lack of identification of electrons or muons, mainly in regions with poor coverage by the electromagnetic calorimetry or muon chambers, or from inefficiencies in the particle identification algorithms, a substantial fraction of \llll\ events was misidentified 
as having a pair of taus. The 48 events selected in the data were classified as follows: 7 in the \eeee~channel, 14 as \eemm, 16 as \eett, 2 as \mmmm, 7 as \mmtt~and 2 as \tttt.

\subsection {$Z \gamma^*$ production in \llll: Results}
\label{sec:llllres}

The value of the $Z\gamma^*$ cross-section at each energy point was extracted using a procedure which followed closely that adopted for the  $l^+ l^- q \bar q$ channels in section~\ref{sec:LLQQRESULTS}.  Bidimensional mass distributions were constructed in the plane of the masses of the pairs with the larger and smaller mass in the event.  The distributions were binned using the same definition as the first five bins in figure~\ref{fig:llqq2} for the  $e^+ e^- q
\bar q$ case. Of the 15.1 events predicted as the $Z\gamma^*$ plus the non-$Z\gamma^*$ \llll\ contributions to the total signal, 3.2 were predicted in the \eeee~channel, 4.5 in \eemm, 3.0 in \eett, 1.6 in \mmmm, 2.2 in \mmtt~and 0.6 in \tttt, while, of the 17 selected data events, 1, 4, 6, 1, 5 and~0 were assigned to each of these channels, respectively.

A one-parameter binned likelihood fit to the $Z \gamma^*$~\llll\ contribution was performed, fixing the non-$Z \gamma^*$ contribution and the remaining backgrounds and interference terms to the Standard Model expectations. These results, shown in the right side of table~\ref{tab:4l1}, were used to derive the combined values of the $Z\gamma^*$ cross-section in the Matrix Element signal definition, as described in section~\ref{sec:RESULTS}.

\subsection{$Z \gamma^*$ production in \llll: Systematic errors}
\label{sec:llllsyst}

Several sources of systematic uncertainties were investigated. 

The main contribution to the systematic error in the track selection came from the difference between data and simulation in the number of reconstructed charged-particle tracks. In order to estimate this uncertainty, samples of dimuon events were generated and the numbers of events with one, two or three reconstructed charged tracks 
compared in data and simulation. From the comparison, a conservative uncertainty of $\pm 5\%$ was assigned as the systematic error from this source. For dimuon events with two reconstructed charged-particle tracks, the difference between data and simulation in the number of events with total charge equal to zero was found to be of the order of +0.5\%.

A contribution of 1.5\% was added due to differences between data and 
simulation in the charge misidentification of electrons in the low polar angle region. 

Systematic uncertainties originating from particle identification were also
taken into account. Two pure samples of $e^+e^-$ and $\mu^+\mu^-$ final states
were selected from the data using particle identification criteria independent 
of those described in section~\ref{sec:PAID} and were compared with 
simulated samples of the same final states. Then the identification criteria 
for electrons and muons were applied to both samples and the difference in the 
efficiencies between data and simulation was taken as a systematic error. 
This resulted in errors of $\pm 0.5\%$ for muons and $\pm 5\%$ for electrons. 
The poorer of the two estimates was also used for taus and adopted as a 
systematic uncertainty on the cross-section measurement.    

Possible errors arising from the procedure adopted in the fits to the $l^+ l^-$
mass distribution were studied. Several checks were performed, in close analogy
to those described in section~\ref{sec:LLQQSYST}. First, simulated samples of
events with electrons in the final state (which receive large contributions 
from $t$-channel processes) were split into two categories, depending on whether or
not the electrons were identified in the event reconstruction. The
cross-sections of the two samples were measured and then combined. Secondly, a
one-parameter fit to the mass distribution was performed, both on the whole
selected \llll\ sample and on the two separated samples with final state
electrons described above, allowing  only the $Z\gamma^*$ component to vary.
From the spread of the results  of these additional fits,  a systematic error 
of $\pm 7\%$ was estimated.

The error in the efficiency for selecting signal events due to the limited 
Monte Carlo statistics was evaluated to be $\pm 0.6\%$. The limited statistics 
available for the different background processes were also taken into account, as well as
the theoretical uncertainties in the cross-sections,  resulting in contributions
of $\pm 0.06\%$ and $\pm 1.1\%$, respectively. Finally, a contribution to the systematic error of $\pm 0.6\%$ was
estimated from the uncertainty in the measurement of the luminosity.

The total estimated systematic error on the \llll\ $Z\gamma^*$ cross-section
measurement was thus estimated to be $\pm 10\%$. 

\subsection{Measurement of the total cross-section for \llll\ production}

In this section we report a total cross-section measurement for \llll\ production, in addition to the study of $Z
\gamma^*$ production in the four-lepton topology described in section~\ref{sec:llllres} above.

As the cross-section does not vary too much within the energy range of LEP2, all the data and the Monte Carlo simulations for the different energies were grouped together. The total cross-section was then estimated from a 
likelihood fit to the Poissonian probability for observing the number of events found in the data, given the expected number corresponding to a total cross-section, $\sigma$, for \llll\ production, plus the estimated number of background events (see table~\ref{tab:4l1}).

The total cross-section for the \llll\ processes was found to be

\begin{center}
$\sigma = (0.430 \pm 0.072 \pm 0.023)$~pb
\end{center}

\noindent within the visible region, defined by $|\cos\theta_l|\leq 0.98$, at  a luminosity-averaged centre-of-mass energy of 197.1~GeV. The first error quoted is statistical; the second is the estimated systematic error, derived as described in section~\ref{sec:llllsyst} above, but without including effects involving particle identification.

This result is in good agreement with the predicted cross-sections  from  WPHACT, which range from 0.440~pb at $\sqrt{s} = 182.7$~GeV to 0.375~pb at $\sqrt{s} = 206.5$~GeV, giving a luminosity-weighted average cross-section of 0.403~pb within the visible region at $\sqrt{s} = 197.1$~GeV.

\section{ Study of the $q \bar q q \bar q$ final state \label{sec:QQQQ}}

The measurement of the $Z \gamma^*$ contribution in the $q \bar q q \bar q$
channel  has to deal with background processes such as $q \bar q (\gamma)$ and
$WW$ which have cross-sections larger by orders of magnitude than the signal. 
It is thus not feasible to measure the $Z \gamma^*$ cross-section in all the
possible $q \bar q$ mass spectrum. Only a restricted  region was thus 
considered here, for low values of the reconstructed mass of one $q \bar q$ pair. The
signature of the process studied in this analysis is the presence of a highly
energetic isolated low mass jet from the $\gamma^*$ hadronisation
(preferentially directed in the forward region), recoiling against a system of
two (or more) jets from the hadronic $Z$ decay. The study of the $\gamma^*$
system was limited to final states with only two charged particles and an
arbitrary number of neutral particles; this choice was driven by the 
expectation that, in the low mass region, the process $\gamma^* \rightarrow q \bar q$ is
dominated by the hadronisation chain  $\gamma^* \rightarrow \rho^0 \rightarrow
\pi^+ \pi^-$. Furthermore, an explicit cut on the reconstructed mass of the 
two selected charged-particle tracks was used, as explained below.  The $Z\gamma^*$ signal
definition was kept the same as in the other channels studied (with no limits 
on the $\gamma^*$ mass); as a result, the two selection criteria  mentioned above
(those requiring low  charged-particle multiplicity and low reconstructed mass) imply a large
inefficiency in the analysis of events with $\gamma^* \rightarrow q \bar q$ for
$\gamma^*$ masses above  a few GeV/$c^2$.

The principal backgrounds arise from production of $q \bar q (\gamma)$, $WW$ 
and final states from other four-fermion neutral current processes such as 
$q \bar q \mu^+ \mu^-$, $q \bar q e^+ e^-$  and $q \bar q \tau^+ \tau^-$.

A pre-selection was applied to the data in order to select hadronic events
compatible with the expected topologies.  The total charged-particle
multiplicity was required  to be larger than 20; the ratio
$\sqrt{s^\prime}/\sqrt s$ had to be larger than 77\%, where $\sqrt{s^\prime}$ is
the reconstructed effective centre-of-mass energy~\cite{sprime}; 
events with neutral particles with electromagnetic energy exceeding 50~GeV were excluded; 
the missing energy of the event was required to be less  than 82\% of the centre-of-mass energy;
and the number of identified muons was required to be less than two (to limit
the background from $q \bar q \mu^+ \mu^-$ events). Events were then clustered
according to the LUCLUS~\cite{luclus}  algorithm with the parameter $d_{join}$
set to 6.5~GeV/$c$, and  it was required that the number of reconstructed jets
in the event be larger than two. One of the jets had to contain at least one
charged particle with momentum exceeding~32 GeV/$c$ and to have 
charged-particle multiplicity of two, while an arbitrary number of neutral particles was accepted in the jet.
The pair of charged particles was then subjected to the selections listed below:

\begin{itemize}
\item The impact parameters of the two charged particles were required to
  be  compatible with production at the primary event vertex;
\item The total energy of the pair was required to be larger than 63~GeV;
\item The two charged particles had to be of opposite charge;
\item The total energy deposited by the two particles in the electromagnetic
calorimeters was required
to be less than 40\% of the total energy of the pair;
\item Identified muons and electrons (soft identification criteria, see 
section~\ref{sec:LLQQ}) were not allowed in the pair;
\item The system recoiling against the jet containing the selected pair was
  forced into a two-jet configuration and the full kinematics of the three
  jets was completely determined by their space directions. Then
  the two-jet system not containing the selected pair was required to have
  a  reconstructed mass within 11~GeV/$c^2$ of the nominal $Z$ mass;
\item The invariant mass of the two charged particles had to be less 
      than 2.1~GeV/$c^2$.
\end{itemize}

\begin{figure}[tbh]
\begin{center}
    \mbox{\epsfysize=13cm\epsffile{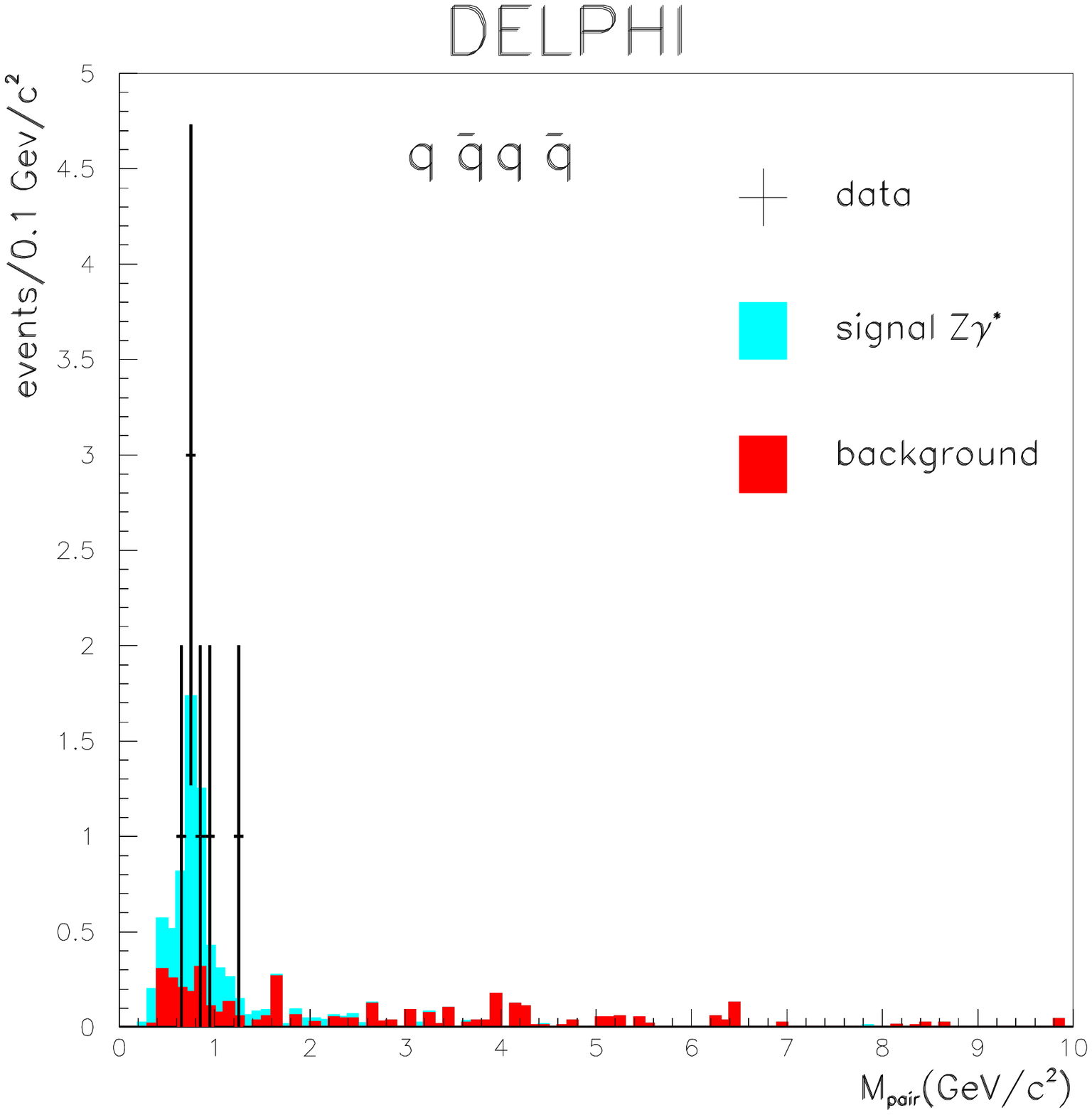}}
\caption{The distribution of the reconstructed invariant mass of the selected
pair of charged-particle tracks in the $q \bar q q \bar q$ analysis, compared with the predictions of the Standard Model. The points are the data, summed over all
energy points and shown before the final selection of $M_{pair} < 2.1$~GeV/$c^2$; the light (blue) histogram shows the predicted $Z\gamma^*$ contribution, and the dark (red) histogram shows the predicted background.}
    \label{fig:massaqqqq}
\end{center}
\end{figure}

Numerical values of the cuts were optimised by scanning the full range of the
relevant discriminating variables and calculating, for each set of values, the
cross-section and the product of the efficiency and purity of the selected 
sample.  The set with the highest value of  the product of efficiency, \( \epsilon \), and purity,
\( p \),  corresponding to \( \epsilon =2.2\% \) and \( p =69.6\% \), was chosen, yielding a 
ratio \( \frac{signal}{\sqrt{background}}=3.3 \). The procedure selected 7 events in data and 6.9 in the simulation, of
which 4.8 were signal and 2.1 were background. The main backgrounds came from
$WW$ (1.1 events), $q \bar q (\gamma)$ (0.4 events) and other
four-fermion neutral current processes (0.4 events). Figure~\ref{fig:massaqqqq}
compares the distribution of the reconstructed mass of the pair of selected
charged-particle tracks before the last cut with Standard Model predictions.
Table~\ref{qqecm} shows the predicted numbers of signal and background events
and the observed numbers of events in the $q \bar q q \bar q$ channel at the
various centre-of-mass energies.

\begin{table}[hbt]
\begin{center}
\begin{tabular}{|c|c|c|c|c|}
\hline
E (GeV) & Data & Total MC &Signal & Background \\ \hline 
182.7  & 1 & 0.4  & 0.4 & 0.1 \\ 
188.6  & 2 & 1.9 & 1.4 & 0.5 \\
191.6  & 0 & 0.2  & 0.2 & 0.0 \\
195.5  & 0 & 0.8  & 0.4 & 0.4  \\
199.5  & 0 & 1.0  & 0.8 & 0.2  \\
201.6  & 2 & 0.3  & 0.2 & 0.0  \\
205.0  & 1 & 0.6  & 0.4 & 0.2  \\
206.5  & 1 & 1.7  & 1.0 & 0.7  \\
\hline
Total  & 7 &6.9 & 4.8  & 2.1 \\ 
\hline
\end{tabular}
\caption []{ 
Observed numbers of events in the $q \bar q q \bar{q}$   channel at each energy
compared with the Standard Model predictions for signal and  background. }
{\label{qqecm}} 
\end{center}
\end{table}

The value of the $Z \gamma^*$ cross-section at each energy point was extracted
using a counting technique and the values were then combined to determine a
global result. All non-$Z \gamma^*$ contributions, backgrounds and interference
terms were fixed to the Standard Model expectations. The results were used to
derive a combined value for the $Z \gamma^*$ cross-section for the Matrix 
Element signal definition, as described in section~\ref{sec:RESULTS}.

\subsection{Systematic errors \label{sec:QQQQSYST}}

Various sources of systematic error were considered.

The predicted background contributions from $WW$, $q \bar q (\gamma)$ and
four-fermion  neutral current production were varied by changing the
cross-sections for these processes according to the values given in
section~\ref{VVQQSYST}: the combined effect on the cross-section 
measurement was estimated to amount to $\pm 0.8\%$.

The statistical error corresponding to the limited simulated sample gave an
uncertainty of $\pm 8\%$.  

The reliability of the simulation in reproducing the amount of background was
checked by repeating the analysis, selecting pairs of particles of the same charge.
The same cuts as those described in section~\ref{sec:QQQQ} were applied, with
the exclusion of the requirement on the total charge of the pair. No events 
were selected in data, while 0.56 were predicted by the simulation. The results are
of course compatible, but to derive a numerical estimate for a systematic error,
the procedure was modified so as to select a larger number of events: the cut 
on the invariant mass of the pair of charged-particle tracks - made at 
2.1~GeV/$c^2$ in the main analysis -  was increased to 10~GeV/$c^2$. All the other 
selections were left unchanged. This gave 3 events
in data and 4.5 in the simulation, of which 3.6 were due to $WW$ production and
0.5 to $q \bar q (\gamma)$ backgrounds. 

A similar study was performed to check the four-fermion neutral current
background, which gave a negligible contribution in the previous procedure. The
selections in section~\ref{sec:QQQQ} were repeated on data and simulation, but
replacing the veto on identified electrons or muons in the selected pair
of charged tracks by the requirement that at least one of the two tracks was
positively identified as a lepton (electron or muon).  In addition, the cut on
the invariant mass of the pair was softened to 10~GeV/$c^2$, as for the check
described in the previous paragraph.  This resulted in 8 events selected in the
data and 6.9 predicted from  the simulation, of which 5.7 were due to the
four-fermion neutral current background (in particular  $l^+ l^- q \bar q$
events, with $l \equiv e,\mu, \tau$) and 0.9 from the $WW$ background. 

As the two last procedures (requirement on the total charge of the pair and on
the presence of leptons in the pair) each showed good agreement between data
and the predictions of the simulation, the results were summed, and the larger
of the statistical error of the data and the difference between data and
simulation was assumed as a systematic uncertainty. This was estimated to be
$\pm 13\%$ on the cross-section measurement.

The uncertainty on the cross-section measurement due to the measurement of the
luminosity was evaluated to be $\pm 0.6\%$.

Finally, the stability of the result as a function of the applied experimental cuts was
checked by varying the numerical values of the analysis selections.  The
procedure set up to maximise the product of the efficiency and purity  
of the simulated sample (see section~\ref{sec:QQQQ}) was used to vary all the relevant
cuts  within reasonable limits; selections were accepted if the predicted 
product of the efficiency and purity of the sample differed by less than the statistical error
of the simulated sample from the optimum value used in the analysis.  For each
new selection, the signal efficiency, background level and number of events in
data were estimated, and a value for the  cross-section was measured. The root
mean square of the distribution of the cross-sections thus obtained was 
evaluated to be 15\% of the central value. As this number is compatible with the statistical fluctuations intrinsic to this procedure, no systematic error was added. 

The total estimated systematic error on the $q \bar q q \bar q$ $Z\gamma^*$
cross-section measurement was thus estimated to be $\pm 15\%$. 

\section{Results \label{sec:RESULTS}}

The measurements described in the previous sections all show good 
agreement with the expectations of the Standard Model. In this section,  
we use these measurements to give  results for the  ratio, $R_{Z \gamma^*}$, 
of the measured to the expected $Z\gamma^*$ cross-section,  for each of the 
final states considered, for their combination at each of the LEP energy 
points at which data were taken, and for the overall average. All these 
results are given in terms of the Matrix Element signal definition 
(see section~\ref{sec:signal}). Results for the LEP signal 
definition are given in section~\ref{sec:RESULTS_LEP}.

\begin{figure}[tbh]
\begin{center}
    \mbox{\epsfysize=13cm\epsffile{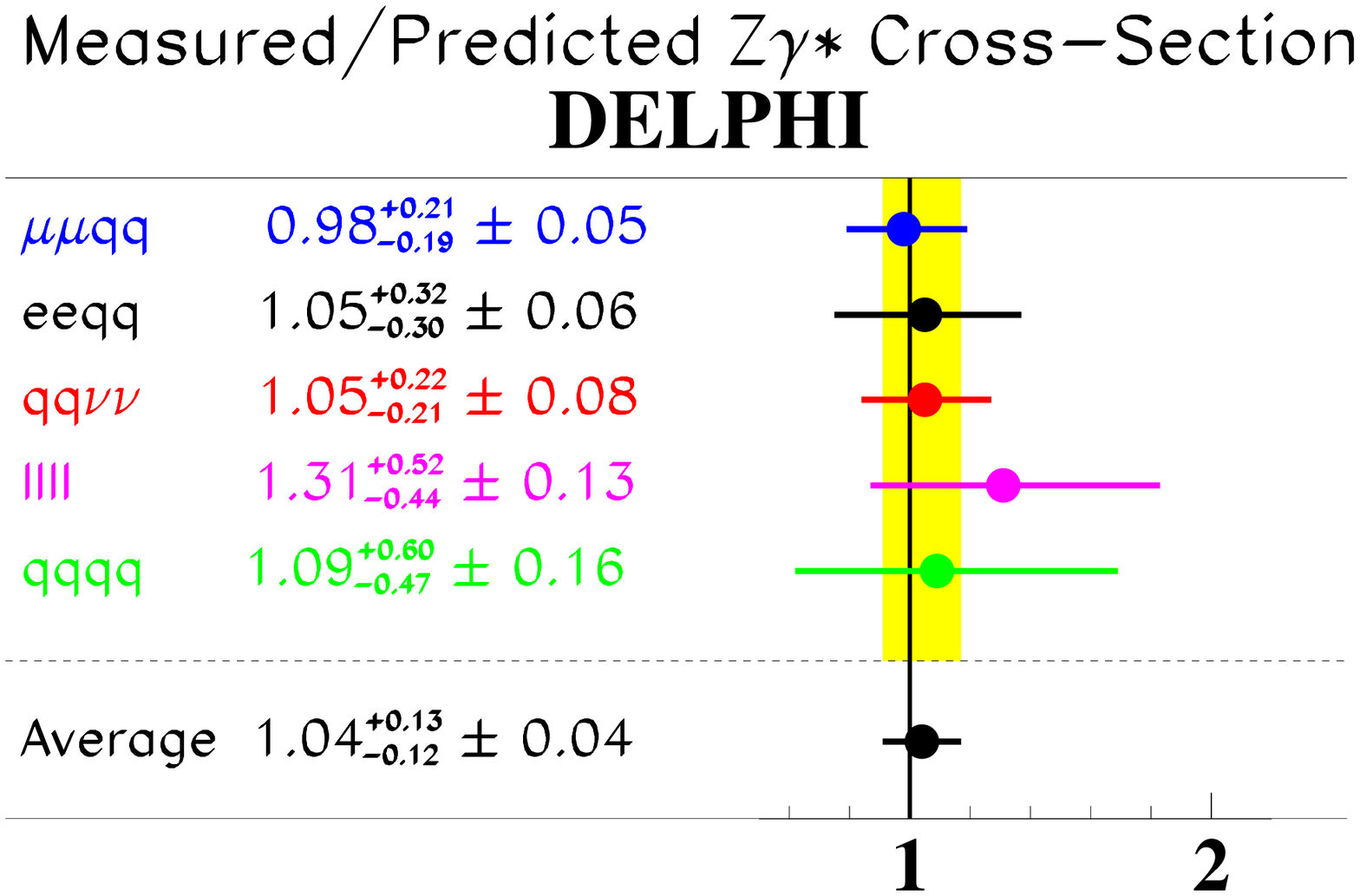}}
\caption{Ratios of measured to predicted cross-sections for individual
  channels contributing to the $Z\gamma^*$ process, using the Matrix Element 
signal definition (see section~\ref{sec:signal}). 
The vertical band displays the total error on  the combination of the 
channels. }
\label{fig:zg_chan}
\end{center}
\end{figure}

Individual cross-sections were extracted by maximising probability functions with respect to the value of the $Z \gamma^*$ cross-section: Poissonian probabilities, based on the number of events selected in data and predicted 
in the simulation, were used for  the $q \bar q \nu \bar \nu$ and $q \bar q q \bar q$ channels; probability functions derived from fitting procedures were used for the $\mu^+ \mu^- q \bar q$, $e^+ e^- q \bar q$ and \llll channels.  
For each centre-of-mass energy, results were expressed in terms of the ratio $R_{Z \gamma^*}$ of measured to expected cross-sections, thus automatically taking into account the  (smooth) dependence with energy predicted by the Standard Model.  The results obtained for the different energies were first combined for each channel separately, and then into a single value. Global likelihoods were  constructed to perform such combinations. The central value was defined as the point of minimum $-\log L$ distribution and the statistical error as the interval around the central  value which contained 68.27\% of the probability. The results obtained for the different channels are shown in table~\ref{tab:results} and in figure~\ref{fig:zg_chan}. The table also shows the average cross-section predicted by the Standard Model for each of the final states considered at the luminosity-weighted average centre-of-mass energy of 197.1~GeV.
 
\begin{figure}[tbh]
\begin{center}
    \mbox{\epsfysize=13cm\epsffile{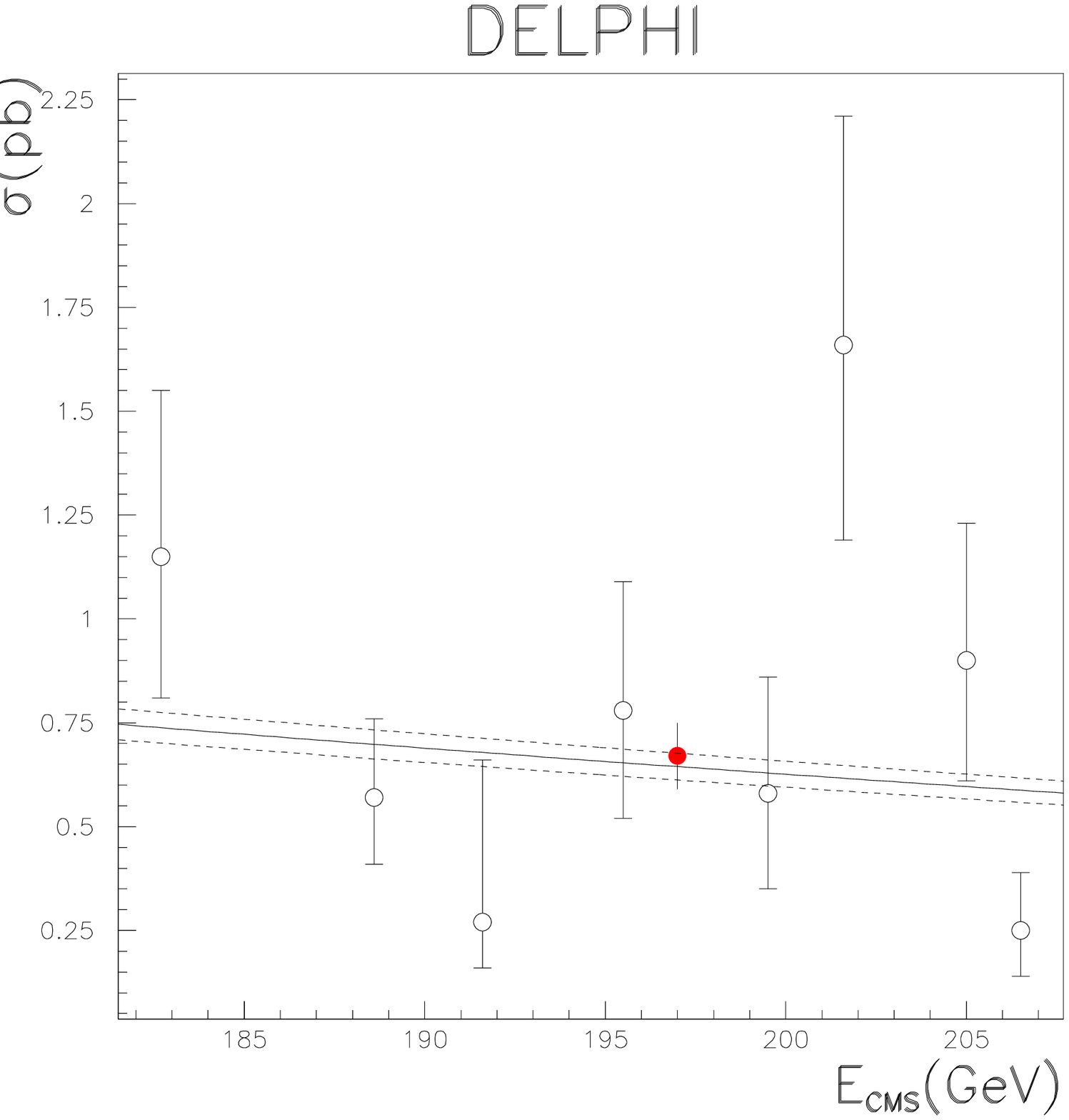}}
\caption{Combined $Z\gamma^*$ cross-section as a function of the 
centre-of-mass energy, using the Matrix Element signal definition 
(see section~\ref{sec:signal}). The solid line is the Standard Model 
prediction; the dashed line represents a 5\%  uncertainty around this 
prediction. The full (red) point is the average cross-section  result, 
plotted at the luminosity-weighted average centre-of-mass energy.}
\label{fig:sigma_ene}
\end{center}
\end{figure}

Table~\ref{tab:res_ene} compares the results at the various energy points, 
averaged over the different channels, with the Standard Model predictions, 
and this comparison is also shown in figure~\ref{fig:sigma_ene}.

The systematic uncertainties for each channel were  studied by introducing appropriately modified assumptions for backgrounds and efficiencies (as described in the corresponding sections) and were considered as fully correlated between the energies.  The effect of systematic uncertainties in the combination of different channels was taken into account considering the uncertainties due to the luminosity measurement and to variations in the predicted background cross-sections as correlated between the channels, and all other effects as uncorrelated.

The final result is  
 $$ R_{Z \gamma^*}~=~1.04~^{+0.13}_{-0.12}(stat)~\pm~0.04(syst)$$
for $ |\cos \theta_{f^\pm}|<0.98$, as shown in tables~\ref{tab:results} 
and~\ref{tab:res_ene} and in figure~\ref{fig:zg_chan}. 
This result is in good agreement with the Standard Model expectation.

\begin{table}[hbt]
\begin{center}
\begin{tabular}{|c|c|c|c|}
\hline
channel             & $R_{Z \gamma^*}$ & $\sigma$ (pb) & $\bar \sigma_{SM} $ (pb)\\ \hline 
\hline
$\mu^+ \mu^- q \bar q$    & $0.98^{+0.21}_{-0.19} \pm 0.05$ &
$0.108^{+0.023}_{-0.021} \pm0.005$ &0.11\\ \hline
$e^+ e^- q \bar q$        & $1.05^{+0.32}_{-0.30} \pm 0.06$ &
$0.115^{+0.035}_{-0.033} \pm 0.007 $ &0.11 \\ \hline
$q \bar q \nu \bar \nu$   & $1.05^{+0.22}_{-0.21} \pm 0.08$ &
$0.084^{+0.018}_{-0.017} \pm 0.006$ &0.08 \\ \hline
\llll                     & $1.31^{+0.52}_{-0.44} \pm 0.13$ &
$0.039^{+0.016}_{-0.013} \pm 0.004$ &0.03 \\ \hline
$q \bar q q \bar q$       & $1.09^{+0.60}_{-0.47} \pm 0.16$ &
$0.316^{+0.174}_{-0.136} \pm 0.047$ &0.29 \\ \hline
\hline
Total                     & $1.04^{+0.13}_{-0.12} \pm 0.04$ &
$0.666^{+0.083}_{-0.077} \pm 0.026$ &0.64\\  \hline
\end{tabular}
\caption []{ 
Ratios of measured to predicted cross-sections and measured cross-sections 
for individual channels contributing to the $Z \gamma^*$ process, using the Matrix Element signal 
definition (see section~\ref{sec:signal}). The first errors are statistical and
the second systematic.  In the last column  $\bar \sigma_{SM}$(pb) is the
average, luminosity-weighted $Z\gamma^*$ cross-section predicted by the
Standard Model at the average energy of 197.1~GeV.}

\label{tab:results}
\end{center}
\end{table}

\begin{table}[hbt]
\begin{center}
\begin{tabular}{|c|c|c|c|}
\hline

E (GeV) & $R_{Z \gamma^*}$ & $\sigma$ (pb)& $\sigma_{SM}$ (pb)\\ \hline \hline
182.7 & 1.55$^{+0.54}_{-0.46} \pm 0.04$ & $1.15^{+0.40}_{-0.34}\pm 0.03 $& 0.74 \\ \hline
188.6 & 0.83$^{+0.27}_{-0.23} \pm 0.04$ & $0.57^{+0.19}_{-0.16}\pm 0.03 $& 0.69 \\ \hline
191.6 & 0.41$^{+0.58}_{-0.17} \pm 0.04$ & $0.27^{+0.39}_{-0.11}\pm 0.03 $& 0.67 \\ \hline
195.5 & 1.18$^{+0.47}_{-0.39} \pm 0.04$ & $0.78^{+0.31}_{-0.26}\pm 0.03 $& 0.66 \\ \hline
199.5 & 0.89$^{+0.43}_{-0.35} \pm 0.04$ & $0.58^{+0.28}_{-0.23}\pm 0.03 $& 0.65 \\ \hline
201.6 & 2.63$^{+0.88}_{-0.74} \pm 0.04$ & $1.66^{+0.55}_{-0.47}\pm 0.03 $& 0.63 \\ \hline
205.0 & 1.52$^{+0.56}_{-0.49} \pm 0.04$ & $0.90^{+0.33}_{-0.29}\pm 0.02 $& 0.59 \\ \hline
206.5 & 0.44$^{+0.24}_{-0.20} \pm 0.04$ & $0.25^{+0.14}_{-0.11}\pm 0.02 $& 0.57 \\ \hline

\hline\hline
Average &   1.04$^{+0.13}_{-0.12} \pm 0.04$ &$0.67^{+0.08}_{-0.08}\pm 0.03 $& 0.64 \\ \hline

\hline
\end{tabular}
\caption []{ Ratios of measured to predicted cross-sections and measured cross-sections averaged over the
different channels at the various energy points, using the Matrix Element signal definition (see section~\ref{sec:signal}). 
The first errors are statistical and the second systematic. The last column shows the Standard Model
predictions, and the entries in the last row refer to the luminosity-averaged centre-of-mass energy of 197.1~GeV. }
\label{tab:res_ene}
\end{center}
\end{table}

\section{Analyses and Results for the LEP signal definition \label{sec:RESULTS_LEP}}

The analyses of the three dominant channels in the final result combination ($\mu^+ \mu^- q \bar q$, $e^+ e^- q \bar q$, $q \bar q \nu \bar \nu$, see table~\ref{tab:results}), described in sections~\ref{sec:LLQQ} and~\ref{sec:VVQQ}, were repeated adopting the LEP signal definition  (see section~\ref{sec:signal}). Some modifications were introduced to the analyses described in the sections referred to above, in order to take into account the fact that the di-fermion
invariant mass regions below the cuts described in section~\ref{sec:signal} must now be considered as background.

\begin{itemize}
\item In the $ l^+ l^- q \bar{q}$ analysis  two additional selections were introduced with respect to those described in
section~\ref{sec:LLQQ}: it was required that the reconstructed mass, $M_{l^+ l^-}$, of the two charged  leptons be larger than 4~GeV/$c^2$ and that the reconstructed mass of the remaining  hadronic system be larger than 8~GeV/$c^2$. This corresponds to reducing the content of bin 1 in the plots of figures~\ref{fig:llqq2},~\ref{fig:llqq3} and~\ref{fig:llqq4} and of bin 4 for muons and bin 6 for electrons in the same figures. The other steps of the analysis  were left unchanged and the same procedures were applied to evaluate the systematic errors.  The total systematic uncertainty on the measured  $Z\gamma^*$ cross-section with the LEP signal definition was estimated to be $\pm 6\%$ for $\mu^+ \mu^-  q \bar q$  and  $\pm 7\%$ for $e^+ e^-  q \bar q$.
\item In the $q \bar q \nu \bar \nu$ analysis similar modifications were introduced.  The low mass analysis (see section~\ref{VVQQ1}) was not used, while in the energy asymmetry and high mass analyses (see sections~\ref{VVQQ2} and~\ref{VVQQ3}, respectively), it was required that the reconstructed mass of the hadronic system be larger than 8~GeV/$c^2$. The other steps of the analyses were left unchanged and the same procedures were applied to evaluate the systematic errors. The total systematic uncertainty  on the measured $Z\gamma^*$ cross-section with the LEP signal definition was estimated to be $\pm 16\%$. 
\end{itemize}

The same procedures as described in section~\ref{sec:RESULTS} were applied in order to obtain  results for the $Z\gamma^*$ cross-sections with the LEP signal definition. The final results for the three channels used are summarised in table~\ref{tab:results_lep}. A combined value of    
\begin{center} 
$ \sigma_{Z\gamma^*}~=~0.136~^{+0.029}_{-0.027}(stat)~\pm~0.008(syst)$ pb
\end{center} 
was obtained for the luminosity-weighted cross-section with the LEP signal definition, in good agreement with the Standard Model prediction of 0.151~pb.

\begin{table}[hbt]
\begin{center}
\begin{tabular}{|c|c|c|c|}
\hline
channel  & $R_{Z \gamma^*}$  & $\sigma$ (pb) & $\bar \sigma_{SM} $ (pb)\\ \hline 
\hline
$\mu^+ \mu^- q \bar q$    &$0.74^{+0.30}_{-0.26} \pm 0.05 $& $0.031^{+0.013}_{-0.011} \pm 0.002$ & 0.042\\ \hline
$e^+ e^- q \bar q$        &$1.05^{+0.30}_{-0.29} \pm 0.08$& $0.061^{+0.017}_{-0.017} \pm 0.004$ & 0.058 \\ \hline
$q \bar q \nu \bar \nu$   &$0.83^{+0.44}_{-0.27} \pm 0.13 $& $0.042^{+0.022}_{-0.014} \pm 0.007$ & 0.051 \\ \hline
\hline
Total                     &$0.90^{+0.19}_{-0.18} \pm 0.05$& 
$0.136 ^{+0.029}_{-0.027} \pm 0.008$ & 0.151\\ \hline
\end{tabular}
\caption []{ 
Ratios of measured to predicted cross-sections and 
luminosity-weighted cross-sections for individual channels contributing to the
$Z \gamma^*$ process, using the LEP signal definition  (see
section~\ref{sec:signal}). The first errors are statistical and the second
systematic.  In the last column  $\bar \sigma_{SM}$(pb) is the average,
luminosity-weighted $Z\gamma^*$ cross-section predicted by the Standard Model.}

\label{tab:results_lep}
\end{center}
\end{table}

\section{Conclusions  \label{sec:CONCL}}

In the data sample collected by the DELPHI detector at centre-of-mass energies ranging from 183~GeV to 209~GeV, the values of the $Z \gamma^*$ cross-section contributing to the four-fermion final states $\mu^+\mu^-q\bar{q}$,
$e^+e^-q\bar{q}$, $q \bar{q} \nu \bar \nu$, $l^+l^-l^+l^-$ and $q \bar q q \bar q$ with $|\cos \theta_{f^\pm}|<0.98$ have been measured and compared with Standard Model expectations. A combined value of 
$$ R_{Z \gamma^*}~=~1.04~^{+0.13}_{-0.12}(stat)~\pm~0.04(syst)$$ 
was obtained for the ratio of the measured to the predicted cross-section in the Matrix Element signal definition (described in section~\ref{sec:signal}). This corresponds to a luminosity-weighted measured cross-section of
\begin{center}
$ \sigma_{Z\gamma^*}~=~0.666~^{+0.083}_{-0.077}(stat)~\pm~0.026(syst)$~pb~,
\end{center}
in good agreement with the value of 0.640~pb predicted by the Standard Model.

Additional cross-section measurements in the channels  $\mu^+\mu^-q\bar{q}$,
$e^+e^-q\bar{q}$ and  $q \bar{q} \nu \bar \nu$ were performed using the
common LEP signal definition (also described in section~\ref{sec:signal}). 
A combined, luminosity-weighted, value of
\begin{center}
 $ \sigma_{Z\gamma^*}~=~0.136~^{+0.029}_{-0.027}(stat)~\pm~0.008(syst)$~pb
\end{center}
was obtained, in good agreement  with the Standard Model prediction of 0.151~pb.

\subsection*{Acknowledgements}
\vskip 3 mm
We are greatly indebted to our technical 
collaborators, to the members of the CERN-SL Division for the excellent 
performance of the LEP collider, and to the funding agencies for their
support in building and operating the DELPHI detector.\\
We acknowledge in particular the support of \\
Austrian Federal Ministry of Education, Science and Culture,
GZ 616.364/2-III/2a/98, \\
FNRS--FWO, Flanders Institute to encourage scientific and technological
research in the industry (IWT) and Belgian Federal Office for Scientific,
Technical and Cultural affairs (OSTC), Belgium, \\
FINEP, CNPq, CAPES, FUJB and FAPERJ, Brazil, \\
Ministry of Education of the Czech Republic, project LC527, \\
Academy of Sciences of the Czech Republic, project AV0Z10100502, \\
Commission of the European Communities (DG XII), \\
Direction des Sciences de la Mati$\grave{\mbox{\rm e}}$re, CEA, France, \\
Bundesministerium f$\ddot{\mbox{\rm u}}$r Bildung, Wissenschaft, Forschung 
und Technologie, Germany,\\
General Secretariat for Research and Technology, Greece, \\
National Science Foundation (NWO) and Foundation for Research on Matter (FOM),
The Netherlands, \\
Norwegian Research Council,  \\
State Committee for Scientific Research, Poland, SPUB-M/CERN/PO3/DZ296/2000,
SPUB-M/CERN/PO3/DZ297/2000, 2P03B 104 19 and 2P03B 69 23(2002-2004)\\
FCT - Funda\c{c}\~ao para a Ci\^encia e Tecnologia, Portugal, \\
Vedecka grantova agentura MS SR, Slovakia, Nr. 95/5195/134, \\
Ministry of Science and Technology of the Republic of Slovenia, \\
CICYT, Spain, AEN99-0950 and AEN99-0761,  \\
The Swedish Research Council,      \\
Particle Physics and Astronomy Research Council, UK, \\
Department of Energy, USA, DE-FG02-01ER41155, \\
EEC RTN contract HPRN-CT-00292-2002. \\


\end{document}